\documentclass[10pt,english,letter]{IEEEtran}
\usepackage{lmodern}

\usepackage[T1]{fontenc}
\usepackage[latin9]{inputenc}
\usepackage{color}
\usepackage{amsthm}
\usepackage{amsmath}
\usepackage{amssymb}
\usepackage{wasysym}
\usepackage{graphicx}

\makeatletter
\theoremstyle{plain}
\newtheorem{thm}{\protect\theoremname}
\theoremstyle{plain}
\newtheorem{prop}[thm]{\protect\propositionname}
\theoremstyle{plain}
\newtheorem{lem}[thm]{\protect\lemmaname}
\theoremstyle{plain}
\newtheorem{cor}[thm]{\protect\corollaryname}

\usepackage{url}
\usepackage{graphics}\usepackage{epsfig}\@ifundefined{definecolor} {\usepackage{color}}{} \usepackage{amsfonts}\usepackage{latexsym}\usepackage{tabularx}

\usepackage{txfonts}

\newcommand{\beq}{\begin{equation}}\newcommand{\eeq}{\end{equation}}\newcommand{\bea}{\begin{eqnarray}}\newcommand{\eea}{\end{eqnarray}}\newcommand{\bda}{\begin{eqnarray*}}\newcommand{\eda}{\end{eqnarray*}}\newcommand{\bdalign}{\begin{align*}}\newcommand{\edalign}{\end{align*}}  

\@ifundefined{showcaptionsetup}{}{%
 \PassOptionsToPackage{caption=false}{subfig}}
\usepackage{subfig}
\makeatother

\usepackage{babel}
\providecommand{\corollaryname}{Corollary}
\providecommand{\lemmaname}{Lemma}
\providecommand{\propositionname}{Proposition}
\providecommand{\theoremname}{Theorem}

\begin{document}

\title{Simple and Effective Dynamic Provisioning for Power-Proportional
Data Centers }

\author{Tan Lu and Minghua Chen\\
Department of Information Engineering, The Chinese University of Hong
Kong}

\maketitle
\begin{abstract}
Energy consumption represents a significant cost in data center operation.
A large fraction of the energy, however, is used to power idle servers
when the workload is low. Dynamic provisioning techniques aim at saving
this portion of the energy, by turning off unnecessary servers. In
this paper, we explore how much performance gain can knowing future
workload information brings to dynamic provisioning. In particular,
we study the dynamic provisioning problem under the cost model that
a running server consumes a fixed amount energy per unit time, and
develop online solutions with and without future workload information
available. We first reveal an elegant structure of the off-line dynamic
provisioning problem, which allows us to characterize and achieve
the optimal solution in a {}``divide-and-conquer'' manner. We then
exploit this insight to design three online algorithms with competitive
ratios $2-\alpha$ , $\left(e-\alpha\right)/\left(e-1\right)\approx1.58-\alpha/\left(e-1\right)$
and $e/\left(e-1+\alpha\right)$, respectively, where $0\leq\alpha\leq1$
is the fraction of a critical window in which future workload information
is available. A fundamental observation is that \emph{future workload
information beyond the critical window will not} \emph{improve dynamic
provisioning performance}. Our algorithms are decentralized and are
simple to implement. We demonstrate their effectiveness in simulations
using real-world traces. We also compare their performance with state-of-the-art
solutions.
\end{abstract}

\section{Introduction}

As Internet services, such as search and social networking, become
more widespread in recent years, the energy consumption of data centers
has been skyrocketing. In 2005, data centers worldwide consumed an
estimated 152 billion kilowatt-hours (kWh) of energy, roughly 1\%
of the world total energy consumption \cite{Koomey2008}. Power consumption
at such level was enough to power half of Italy \cite{worldenergy07}.
Energy cost is approaching overall hardware cost in data centers \cite{barroso2005price},
and is growing 12\% annually \cite{ESPreport2007}.

Recent works have explored electricity price fluctuation in time and
geographically load balancing across data centers to cut short the
electricity bill; see e.g., \cite{liu2011greening,wendell2010donar,qureshi2009cutting,urgaonkar2011optimal}
and the references therein. Meanwhile, it is nevertheless critical
to minimize the actual energy footprint in individual data centers.

Energy consumption in a data center is a product of the PUE%
\footnote{Power usage effectiveness (PUE) is defined as the ratio between the
amount of power entering a data center and the power used to run its
computer infrastructure. The closer to one PUE is, the better energy
utilization is.%
} and the energy consumed by the servers. There have been substantial
efforts in improving PUE, e.g., by optimizing cooling \cite{rasmussen113electrical,sharma2005balance}
and power management \cite{raghavendra2008no}. We focus on reducing
the energy consumed by the servers in this paper.

Real-world statistics reveals three observations that suggest ample
saving is possible in server energy consumption \cite{chase2001managing,pinheiro2001load,chen2008energy,krioukov2011napsac,fan2007power,barroso2007case}.
First, workload in a data center often fluctuates significantly on
the timescale of hours or days, expressing a large {}``peak-to-mean''
ratio. Second, data centers today often provision for far more than
the observed peak to accommodate both the predictable workload and
the unpredictable flash crowds%
\footnote{In May 2011, Amazon\textquoteright{}s data center is down for hours
due to a surge downloads of Lady Gaga's song {}``Born This Way''.%
}. Such static over-provisioning results in low average utilization
for most servers in data centers. Third, a low-utilized or idle server
consumes more than 60\% of its peak power. These observations imply
that a large portion of the energy consumed by servers goes into powering
nearly-idle servers, and it can be best saved by turning off servers
during the off-peak periods.

One promising technique exploiting the above insights is \emph{dynamic
provisioning}, which turns on a minimum number of servers to meet
the current demand and dispatches the load among the running servers
to meet Service Level Agreements (SLA), making the data center {}``power-proportional''.

There have been a significant amount of efforts in developing such
technique, initiated by the pioneering works \cite{chase2001managing}\cite{pinheiro2001load}
a decade ago. Among them, one line of works \cite{meisner2009powernap,krioukov2011napsac,chen2008energy}
exam the practical feasibility and advantage of dynamic provisioning
using real-world traces, suggesting substantial gain is indeed possible
in practice. Another line of works \cite{chase2001managing,qian2011server,lin2011dynamic,chen2008energy}
focus on developing algorithms by utilizing various tools from queuing
theory, control theory, and machine learning, providing algorithmic
insights in synthesizing effective solutions. These existing works
provide a number of schemes that deliver favorable performance justified
by theoretic analysis and/or practical evaluations. See \cite{DataCenterEnergySurvey10}
for a recent survey.

The effectiveness of these exciting schemes, however, usually rely
on being able to predict future workload to certain extent, e.g.,
using model fitting to forecast future workload from historical data
\cite{chen2008energy}. This naturally leads to the following questions:
\begin{itemize}
\item Can we design \emph{online} solutions that require zero future workload
information, yet still achieve \emph{close-to-optimal} performance?
\item Can we characterize the benefit of knowing future workload in dynamic
provisioning?
\end{itemize}
Answers to these questions provide fundamental understanding on how
much performance gain one can have by exploiting future workload information
in dynamic provisioning.

Recently, Lin \emph{et al.} \cite{lin2011dynamic} propose an algorithm
that requires almost-zero future workload information%
\footnote{The LCP algorithm proposed in \cite{lin2011dynamic} only relies on
an estimate of the job arrival rate of the upcoming slot.%
} and achieves a competitive ratio of 3, i.e., the energy consumption
is at most 3 times the minimum (computed with perfect future knowledge).
In simulations, they further show the algorithm can exploit available
future workload information to improve the performance. These results
are very encouraging, indicating that a complete answer to the questions
is possible.

In this paper, we further explore answers to the questions, and make
the following contributions:
\begin{itemize}
\item We consider a scenario where a running server consumes a fixed amount
energy per unit time. We reveal that the dynamic provisioning problem
has an elegant structure that allows us to solve it in a {}``divide-and-conquer''
manner. This insight leads to a full characterization of the optimal
solution, achieved by using a centralized procedure.
\item We show that, interestingly, the optimal solution can also be attained
by the data center adopting a simple \emph{last-empty-server-first}
job-dispatching strategy%
\footnote{Readers might notice that this job-dispatching strategy shares some
similarity with the most-recently-busy strategy used in the DELAYEDOFF
algorithm \cite{gandhi2010optimality}. Actually there are subtle
yet important difference, which will be discussed in details in Section
\ref{ssec:comparison.with.DELAYEDOFF}.%
} and each server \emph{independently} solving a classic ski-rental
problem. We build upon this architectural insight to design three
\emph{decentralized} online algorithms, all have improved competitive
ratios than state-of-the-art solutions. One is a deterministic algorithm
with competitive ratio $2-\alpha$, where $0\leq\alpha\leq1$ is the
fraction of a critical window in which future workload information
is available. The other two are randomized algorithms with competitive
ratios $\left(e-\alpha\right)/\left(e-1\right)\approx1.58-\alpha/\left(e-1\right)$and
$e/\left(e-1+\alpha\right)$, respectively. We prove that $2-\alpha$
and $e/\left(e-1+\alpha\right)$ are the best competitive ratios for
deterministic and randomized online algorithms under our last-empty-server-first
job-dispatching strategy.
\item Our results lead to a fundamental observation: under the cost model
that a running server consumes a fixed amount energy per unit time,
\emph{future workload information beyond the critical window will
not} \emph{improve the dynamic provisioning performance. }The size
of the critical window is determined by the wear-and-tear cost and
the unit-time energy cost of running one server.
\item Our algorithms are simple and easy to implement. We demonstrate the
effectiveness of our algorithms in simulations using real-world traces.
We also compare their performance with state-of-the-art solutions.
\end{itemize}

The rest of the paper is organized as follows. We formulate the problem
in Section \ref{sec:ps}. Section \ref{sec:offline} reveals the important
structure of the formulated problem, characterizes the optimal solution,
and designs a simple decentralized offline algorithm achieving the
optimal. In Section \ref{sec:online}, we propose the online algorithms
and provide performance guarantees. Section \ref{sec:expr} presents
the numerical experiments and Section \ref{sec:conclusion} concludes
the paper.

\section{Problem Formulation\label{sec:ps}}

\subsection{Settings and Models \label{ssec:settings}}

We consider a data center consisting of a set of homogeneous servers.
Without loss of generality, we assume each server has a unit service
capacity%
\footnote{In practice, server's service capacity can be determined from the
knee of its throughput and response-time curve \cite{krioukov2011napsac}.%
}, i.e., it can only serve one unit workload per unit time. Each server
consumes $P$ energy per unit time if it is on and zero otherwise.
We define $\beta_{on}$ and $\beta_{off}$ as the cost of turning
a server on and off, respectively. Such wear-and-tear cost, including
the amortized service interruption and hard-disk failure cost\cite{qian2011server},
is comparable to the energy cost of running a server for several hours
\cite{lin2011dynamic}.

The results we develop in this paper apply to both of the following
two types of workload%
\footnote{There are also other types of workload, such as the bin-packing model
considered in \cite{krioukov2011napsac}. Extending the results in
this paper to those workload models is of great interest and left
for future work.%
}:
\begin{itemize}
\item {}``mice'' type of workload, such as {}``request-response'' web
serving. Each job of this type has a small transaction size and short
duration. A number of existing works \cite{chase2001managing,pinheiro2001load,lin2011dynamic,doyle2003model}
model such workload by a discrete-time fluid model. In the model,
time is chopped into equal-length slots. Jobs arriving in one slot
get served in the same slot. Workload can be split among running servers
at arbitrary granularity like fluid.
\item {}``elephant'' type of workload, such as virtual machine hosting
in cloud computing. Each job of this type has a large transaction
size, and can last for a long time. We model such workload by a continuous-time
brick model. In this model, time is continuous, and we assume one
server can only serve one job%
\footnote{Other than the obvious reason that the service capacity can only fit
one job, there could also be SLA in cloud computing that requires
the job does not share the physical server with other jobs due to
security concerns.%
}. Jobs arrive and depart at arbitrary time, and no two job arrival/departure
events happen simultaneously.
\end{itemize}
For the discrete-time fluid model, servers toggled at the discrete
time epoch will not interrupt job execution and thus no job migration
is incurred. This neat abstraction allows research to focus on server
on-off scheduling to minimize the cost. For the continuous-time brick
model, when a server is turned off, the long-lasting job running on
it needs to be migrated to another server. In general, such non-trivial
migration cost needs to be taken into account when toggling servers.

In the following, we present our results based on the continuous-time
brick model. We add discussions to show the algorithms and results
are also applicable to the discrete-time fluid model.

Let $x\left(t\right)$ and $a\left(t\right)$ be the number of {}``on''
servers (serving or idle) and jobs at time $t$ in the data center,
respectively. To keep the problem interesting, we assume that $a\left(t\right)$
is not always zero. Under our workload model, $a(t)$ at most increases
or decreases by one at any time $t$.

To focus on the cost within $[0,T]$, we set $x(0)=a\left(0\right)$
and $x\left(T\right)=a\left(T\right)$. Note such boundary conditions
include the one considered in the literature, e.g., \cite{lin2011dynamic},
as a special case, where $x(0)=a(0)=x(T)=a(T)=0$.

Let $P_{on}(t_{1},t_{2})$ and $P_{off}(t_{1},t_{2})$ denote the
total wear-and-tear cost incurred by turning on and off servers in
$[t_{1},t_{2}]$, respectively:
\begin{equation}
P_{on}(t_{1},t_{2})\triangleq\underset{\delta\rightarrow0^{+}}{\lim}\left\{ \beta_{on}\underset{i=1}{\overset{\left\lceil \left(t_{2}-t_{1}\right)/\delta\right\rceil }{\sum}}\left[x\left(t_{1}+i\delta\right)-x\left(t_{1}+\left(i-1\right)\delta\right)\right]^{+}\right\} \label{eq:on-cost}
\end{equation}
and
\begin{equation}
P_{off}(t_{1},t_{2})\triangleq\underset{\delta\rightarrow0^{+}}{\lim}\left\{ \beta_{off}\underset{i=1}{\overset{\left\lceil \left(t_{2}-t_{1}\right)/\delta\right\rceil }{\sum}}\left[x\left(t_{1}+\left(i-1\right)\delta\right)-x\left(t_{1}+i\delta\right)\right]^{+}\right\} .\label{eq:off-cost}
\end{equation}

\subsection{Problem Formulation}

We formulate the problem of minimizing server operation cost in a
data center in $[0,T]$ as follows:
\begin{eqnarray}
\mathbf{SCP}: & \textrm{min} & P\varint_{0}^{T}x\left(t\right)dt+P_{on}(0,T)+P_{off}(0,T)\label{eq: obj}\\
 & \textrm{s.t}. & x(t)\geq a(t),\forall t\in[0,T],\label{eq:const1}\\
 &  & x(0)=a(0),x(T)=a(T),\label{eq:asym.constraint}\\
 & \mbox{var} & x(t)\in\mathbb{Z}^{+},t\in[0,T],
\end{eqnarray}
where $\mathbb{Z}^{+}$ denotes the set of non-negative integers.

The objective is to minimize the sum of server energy consumption
and the wear-and-tear cost. Constraints in \eqref{eq:const1} say
the service capacity must satisfy the demand. Constraints in \eqref{eq:asym.constraint}
are the boundary conditions.

\textbf{Remarks}: (i) The problem \textbf{SCP} does not consider the
possible migration cost associated with the continuous-time discrete-load
model. Fortunately, our results later show that we can schedule servers
according to the optimal solution, and at the same time dispatch jobs
to servers in a way that aligns with their on-off schedules, thus
incurring no migration cost. Hence, the minimum server operation cost
remains unaltered even we consider migration cost in the problem \textbf{SCP}
(which can be rather complicated to model). (ii) The formulation remains
the same with discrete-time fluid workload model where there is no
job migration cost to consider. (iii) The problem\textbf{ SCP} is
similar to a common one considered in the literature, e.g., in \cite{lin2011dynamic},
with a specific cost function. The difference is that we allow more
flexible boundary conditions and on/off wear-and-tear cost modeling,
and are more precise in the decision variables being integers instead
of real numbers.(iv) In the problem setting, we assume that the power
consumption of a server is constant $P.$ Actually, the results of
this paper also apply to the following unit time power consumption
model: the power consumption of $x$ busy server is $F\left(x\right)$
and the unit time power consumption for a idle server is $P$. This
is because the total power consumption under this model is $\varint_{0}^{T}F\left[a\left(t\right)\right]+P\left[x\left(t\right)-a\left(t\right)\right]dt+P_{on}(0,T)+P_{off}(0,T)$.
Since $\varint_{0}^{T}F\left[a\left(t\right)\right]-Pa\left(t\right)dt$
is constant for given $a\left(t\right)$, to minimize the total power
consumption is to minimize above \textbf{SCP} problem.

There are infinite number of integer variables $x\left(t\right)$,
$t\in[0,T]$, in the problem \textbf{SCP}, which make it challenging
to solve. Moreover, in practice the data center has to solve the problem
without knowing the workload $a(t)$, $t\in[0,T]$ ahead of time.

Next, we first focus on designing off-line solution, including (i)
a job-dispatching algorithm and (ii) a server on-off scheduling algorithm,
to solve the problem \textbf{SCP} optimally. We then extend the solution
to its on-line versions and analyze their performance guarantees with
or without (partial) future workload information.

\section{Optimal Solution and Offline Algorithm \label{sec:offline}}

We study the off-line version of the server cost minimization problem
\textbf{SCP}, where the workload $a(t)$ in $[0,T]$ is given.

We first identify an elegant structure of its optimal solution, which
allows us to solve the problem in a {}``divide-and-conquer'' manner.
That is, to solve the problem \textbf{SCP} in $[0,T]$, it suffices
to split it into smaller problems over certain \emph{critical segments}
and solve them independently. We then derive a simple and decentralized
algorithm, upon which we build our online algorithms.

\subsection{Critical Times and Critical Segments}

Given $a(t)$ in $[0,T]$, we identify a set of critical times $\left\{ T_{i}^{c}\right\} _{i}$
and construct the \emph{critical segments} as follows. \\
\rule{1\columnwidth}{1pt}

\noindent \textbf{Critical Segment Construction Procedure: }

\begin{figure}
\centering\includegraphics[width=0.9\columnwidth]{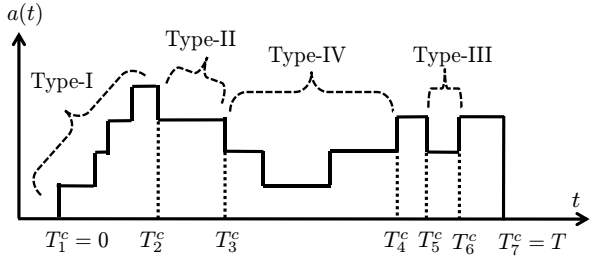}

\caption{Illustration of critical times and critical segments. $T_{1}^{c}$
to $T_{7}^{c}$ are critical times, and they form six critical segments.
$a(t)$ is of Type-I in $\left[T_{1}^{c},T_{2}^{c}\right]$, Type-II
in $\left[T_{2}^{c},T_{3}^{c}\right]$, Type-III in $\left[T_{5}^{c},T_{6}^{c}\right]$,
and Type-IV in $\left[T_{3}^{c},T_{4}^{c}\right]$.}
\label{fig:ct.cs.example}

\end{figure}

First, traversing $a(t)$, we identify all the jobs arrival/departure
epochs in $[0,T]$. The first critical time is $T_{1}^{c}=0$. $T_{1}^{c}$
can be a job-arrival epoch or job-departure epoch, or no job departs/arrive
the system at $T_{1}^{c}$. If no job departs or arrives at $T_{1}^{c}$,
$T_{1}^{c}$ is considered as a job-arrival epoch. Next we find $T_{i+1}^{c}$
inductively, given that $T_{i}^{c}$ is known.
\begin{itemize}
\item If $T_{i}^{c}$ is a job-arrival epoch, e.g., the first critical time,
then $T_{i+1}^{c}$ is the first job-departure epoch after $T_{i}^{c}$.
One example is the epoch $T_{2}^{c}$ in Fig. \ref{fig:ct.cs.example}.
\item If $T_{i}^{c}$ is a job-departure epoch, we first try to find the
first arrival epoch $\tau$ after $T_{i}^{c}$ so that $a\left(\tau\right)=a\left(T_{i}^{c}\right)$.
If such $\tau$ exists, then we set $T_{i+1}^{c}=\tau$. One example
is the epoch $T_{4}^{c}$ in Fig. \ref{fig:ct.cs.example}. If no
such $\tau$ exists, and we set $T_{i+1}^{c}$ to be the next job
departure epoch. One example is the $T_{3}^{c}$ in Fig. \ref{fig:ct.cs.example}.
\end{itemize}
Upon reaching time epoch $T$, we find all, say $M$, critical times.
We define the critical segments as the period between two consecutive
critical times, i.e., $\left[T_{i}^{c},T_{i+1}^{c}\right]$, $1\leq i\leq M-1$.
\\
\rule[0.5ex]{1\columnwidth}{1pt}

The critical segments have interesting properties. For example, they
are disjoint except at the boundary points, and they together fully
cover the time interval $[0,T]$. Moreover, we observe that workload
expresses interesting properties in these critical segments.

\begin{prop}
The workload $a(t)$ in any critical segment \textup{$\left[T_{i}^{c},T_{i+1}^{c}\right]$}
must be one of the following four types\textup{:\label{prop:property}}
\begin{itemize}
\item Type-I: \textup{workload is non-decreasing in $\left[T_{i}^{c},T_{i+1}^{c}\right]$.
\label{enu:property 1}}
\item Type-II: \textup{workload is step-decreasing in $\left[T_{i}^{c},T_{i+1}^{c}\right]$.
That is, $a\left(t\right)=a\left(T_{i}^{c}\right)-1,\forall t\in\left(T_{i}^{c},T_{i+1}^{c}\right]$
and $a\left(t\right)\leq a\left(T_{i}^{c}\right)-1,\forall t\in\left(T_{i+1}^{c},T\right]$.
\label{enu:property 2}}
\item Type-III: \textup{workload is of {}``U-shape'' in $\left[T_{i}^{c},T_{i+1}^{c}\right]$.
That is, $a\left(T_{i+1}^{c}\right)=a\left(T_{i}^{c}\right)$ and
$a\left(t\right)=a\left(T_{i}^{c}\right)-1,\forall t\in\left(T_{i}^{c},T_{i+1}^{c}\right)$.
\label{enu:property 3}}
\item Type-IV: \textup{workload is of {}``canyon-shape'' in $\left[T_{i}^{c},T_{i+1}^{c}\right]$.
That is, $a\left(T_{i+1}^{c}\right)=a\left(T_{i}^{c}\right)$, $a\left(t\right)\leq a\left(T_{i}^{c}\right)-1$
and not always identical, $\forall t\in\left(T_{i}^{c},T_{i+1}^{c}\right)$.
\label{enu:property 4}}
\end{itemize}
\end{prop}
\begin{IEEEproof}
Refer to Appendix \ref{apx:proof_1}.
\end{IEEEproof}
Examples of these four types of $a(t)$ are shown in Fig. \ref{fig:ct.cs.example}.

\subsection{Structure of Optimal Solution }

Let $x^{*}(t)$, $t\in[0,T]$, be an optimal solution to the problem
\textbf{SCP}, and the corresponding minimum server operation cost
be $P^{*}$. We have the following observation.
\begin{lem}
$x^{*}\left(t\right)$ must meet $a\left(t\right)$ at every critical
time, i.e., $x^{*}\left(T_{i}^{c}\right)=a\left(T_{i}^{c}\right)$,
$1\leq i\leq M$.\label{lem:lemma 2}\end{lem}
\begin{IEEEproof}
Refer to Appendix \ref{apx:proof_2}.
\end{IEEEproof}
Lemma \ref{lem:lemma 2} not only presents a necessary condition for
a solution $x(t)$ to be optimal, but also suggests a {}``divide-and-conquer''
way to solve the problem \textbf{SCP} optimally.

Consider the following sub-problem of minimizing server operation
cost in a critical segment $\left[T_{i}^{c},T_{i+1}^{c}\right]$,
$1\leq i\leq M-1$:
\begin{eqnarray}
 & \textrm{min} & P\varint_{T_{i}^{c}}^{T_{i+1}^{c}}x\left(t\right)dt+P_{on}\left(T_{i}^{c},T_{i+1}^{c}\right)+P_{off}\left(T_{i}^{c},T_{i+1}^{c}\right)\label{eq:sub}\\
 & \textrm{s.t}. & x(t)\geq a(t),\forall t\in\left[T_{i}^{c},T_{i+1}^{c}\right],\\
 &  & x(T_{i}^{c})=a(T_{i}^{c}),x(T_{i+1}^{c})=a(T_{i+1}^{c}),\label{eq:sub-const}\\
 & \mbox{var} & x(t)\in\mathbb{Z}^{+},t\in\left[T_{i}^{c},T_{i+1}^{c}\right].
\end{eqnarray}
Let its optimal value be $P_{i}^{*}$, $1\leq i\leq M-1$. We have
the following observation.
\begin{lem}
\label{prop: lower bound}$\underset{i=1}{\overset{M}{\sum}}P_{i}^{*}$
is a lower bound of the optimal server operation cost of the problem
\textbf{SCP}, i.e.,
\begin{equation}
P^{*}\geq\underset{i=1}{\overset{M}{\sum}}P_{i}^{*}.\label{eq:P_opt_lower_bound}
\end{equation}
\end{lem}
\begin{IEEEproof}
Refer to Appendix \ref{apx:proof_3}.
\end{IEEEproof}
\textbf{Remark}: Over arbitrarily chopped segments, sum of their minimum
server operation costs may not be bounds for $P^{*}$. However, as
we will see later, computed based on critical segments, Eqn. \eqref{eq:P_opt_lower_bound}
establishes a lower bound of $P^{*}$ and is achievable, thanks to
the structure of $x^{*}\left(t\right)$ outlined in Lemma \ref{lem:lemma 2}.

Suggested by Lemma \ref{prop: lower bound}, it suffices to solve
individual sub-problems for all critical segments in $[0,T]$, and
combine the corresponding solutions to form an optimal solution to
the overall problem \textbf{SCP} (note the optimal solutions of sub-problems
connect seamlessly). The special structures of $a(t)$ in individual
critical segment, summarized in Proposition \ref{prop:property},
are the key to tackle each sub-problem. \rule{1\columnwidth}{1pt}\\
\textbf{Optimal Solution} \textbf{Construction Procedure}:

We visit all the critical segments in $[0,T]$ sequentially, and construct
an $x(t)$, $t\in[0,T]$. For a critical segment $\left[T_{i}^{c},T_{i+1}^{c}\right]$,
$1\leq i\leq M-1$, we check the $a(t)$ in it:
\begin{enumerate}
\item the $a(t)$ is of Type-I or Type-II: we simply set $x(t)=a(t)$, for
all $t\in$$\left[T_{i}^{c},T_{i+1}^{c}\right]$.
\item the $a(t)$ is of Type-III:

\begin{itemize}
\item if $\beta_{on}+\beta_{off}\geq P\cdot\left(T_{i+1}^{c}-T_{i}^{c}\right)$,
then we set $x\left(t\right)=a\left(T_{i}^{c}\right),\forall t\in\left[T_{i}^{c},T_{i+1}^{c}\right]$;
\item otherwise, we set $x(T_{i}^{c})=a(T_{i}^{c})$, $x(T_{i+1}^{c})=a(T_{i+1}^{c})$,
and $x\left(t\right)=a\left(T_{i}^{c}\right)-1,\forall t\in\left(T_{i}^{c},T_{i+1}^{c}\right)$.
\end{itemize}
\item the $a(t)$ is of Type-IV:

\begin{itemize}
\item if $\beta_{on}+\beta_{off}\geq P\cdot\left(T_{i+1}^{c}-T_{i}^{c}\right)$,
then we set $x\left(t\right)=a\left(T_{i-1}^{c}\right),\forall t\in\left[T_{i}^{c},T_{i+1}^{c}\right]$;
\item Otherwise, we construct $x\left(t\right)$ as follows. In Type-IV
critical segment, each job-departure epoch $\tau$ in $\left[T_{i}^{c},T_{i+1}^{c}\right]$
has a corresponding job-arrival epoch $\tau^{'}$ in $\left[T_{i}^{c},T_{i+1}^{c}\right]$
such that $a\left(\tau\right)=a\left(\tau^{'}\right)$ and $a\left(t\right)<a\left(\tau\right),\forall t\in\left(\tau,\tau^{'}\right)$.
Finding the first job-departure epoch $\tau_{1}$ after $T_{i}^{c}$
in $\left[T_{i}^{c},T_{i+1}^{c}\right]$ who has a corresponding job-arrival
epoch $\tau_{1}^{'}$ such that $\beta_{on}+\beta_{off}\geq P\cdot\left(\tau_{1}^{'}-\tau_{1}\right)$.
Then finding the first job-departure epoch $\tau_{2}$ after $\tau_{1}^{'}$
who has a corresponding job-arrival epoch $\tau_{2}^{'}$ such that
$\beta_{on}+\beta_{off}\geq P\cdot\left(\tau_{2}^{'}-\tau_{2}\right)$.
Go on this way until we reach $T_{i+1}^{c}$. Upon reaching time epoch
$T_{i+1}^{c}$, we find all, say $L$, such job-departure and arrival
epoch pairs $\left(\tau_{1},\tau_{1}^{'}\right)$,$\left(\tau_{2},\tau_{2}^{'}\right)$...$\left(\tau_{L},\tau_{L}^{'}\right)$.
If $L=0$, which means there does not exist such job-departure and
arrival epoch pair, we set $x\left(t\right)=a\left(t\right),\forall t\in\left[T_{i}^{c},T_{i+1}^{c}\right]$,
otherwise, we set $x\left(t\right)=a\left(t\right),\forall t\in\left[T_{i}^{c},\tau_{1}\right)\cup\left(\tau_{1}^{'},\tau_{2}\right)\cup...\cup\left(\tau_{L}^{'},T_{i+1}^{c}\right]$
and $x\left(t\right)=a\left(\tau_{l}\right),\forall t\in\left[\tau_{l},\tau_{l}^{'}\right]$
for $l=1,2,....L$.
\end{itemize}
\end{enumerate}
\rule{1\columnwidth}{1pt}

The following theorem shows that the lower bound of $P^{*}$ in \eqref{eq:P_opt_lower_bound}
is achieved by using the above procedure.
\begin{thm}
The \textbf{Optimal Solution} \textbf{Construction Procedure} terminates
in finite time, and the resulting $x\left(t\right)$, $t\in[0,T]$,
is an optimal solution to the problem \textbf{SCP}.\label{Thm:opt_sol_pro_is_opt}\end{thm}
\begin{IEEEproof}
Refer to Appendix \ref{apx:proof_4}.
\end{IEEEproof}
The proof utilizes proof-by-contradiction and counting arguments.

\subsection{Intuitions and Observations}

\begin{figure}
\centering\includegraphics[width=0.65\columnwidth]{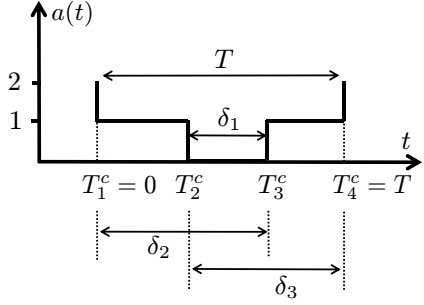}

\caption{An example of a critical segment $[0,T]$ (after offsetting the time
origin to the beginning of the segment) with Type-IV $a(t)$. This
critical segment is further decomposed into smaller critical segments
$[T_{1}^{c},T_{2}^{c}]$, $[T_{2}^{c},T_{3}^{c}]$, and $[T_{3}^{c},T_{4}^{c}]$.
Interval $\delta_{1}=T_{3}^{c}-T_{2}^{c}$, $\delta_{2}=T_{3}^{c}-T_{1}^{c}$,
and $\delta_{3}=T_{4}^{c}-T_{2}^{c}$.}
\label{fig:optimal_type_IV}
\end{figure}

Constructing optimal $x(t)$ for critical segments with Type-I/II/III
workload is rather straightforward. In the following, we go through
the construction of $x(t)$ for the critical segment with Type-IV
workload shown in Fig. \ref{fig:optimal_type_IV}, to bring out the
intuition. We define
\begin{equation}
\Delta\triangleq\frac{\beta_{on}+\beta_{off}}{P}\label{eq:critical_interval}
\end{equation}
as the \emph{critical interval} over which the energy cost of maintaining
an idle server matches the cost of turning it off at the beginning
of the interval and turning it on at the end of the interval.

During the critical segment $[0,T]$ with Type-IV workload shown in
Fig. \ref{fig:optimal_type_IV}, the system starts and ends with 2
jobs and 2 running servers. Let the servers with their jobs leaving
at time $0$ and $T_{3}^{c}$ be S1 and S2, respectively.

At time $0$, a job leaves. The procedure compares $\Delta$ and $T$.
If $\Delta>T$, then it sets $x(t)=2$ and keeps all two servers running
for all $t\in[0,T]$; otherwise, it further applies the \textbf{Critical
Segment Construction Procedure} and decomposes the critical segment
into three small ones $[T_{1}^{c},T_{2}^{c}]$, $[T_{2}^{c},T_{3}^{c}]$,
and $[T_{3}^{c},T_{4}^{c}]$, as shown in Fig. \ref{fig:optimal_type_IV}.
The first small critical segment $[T_{1}^{c},T_{2}^{c}]$ has a Type-II
workload, thus the procedure sets $x(t)=1$ for $t\in[T_{1}^{c},T_{2}^{c}]$.
The second small segment $[T_{2}^{c},T_{3}^{c}]$ has a Type-III workload;
thus for all $t\in[T_{2}^{c},T_{3}^{c}]$, the procedure maintains
$x(t)=1$ if $\Delta>\delta_{1}$ and sets $x(t)=0$ otherwise. The
last small segment $[T_{3}^{c},T_{4}^{c}]$ has a Type-I workload,
thus the procedure set $x(t)=1$ for $t\in[T_{3}^{c},T_{4}^{c})$
and $x(T_{4}^{c})=2$.

These actions reveal two important observations, upon which we build
a decentralized off-line algorithm to solve the problem \textbf{SCP}
optimally.
\begin{itemize}
\item Newly arrived jobs should be assigned to servers in the reverse order
of their last-empty-epochs.
\end{itemize}
In the example, when a new job arrives at time $T_{3}^{c}$, the procedure
implicitly assigns it to server S2 instead of S1. As a result, S1
and S2 have empty periods of $T$ and $\delta_{1}$, respectively.
This may sound counter-intuitive as compared to an alternative {}``fair''
strategy that assigns the job to the early-emptied server S1, which
gives S1 and S2 empty periods of $\delta_{2}$ and $\delta_{3}$,
respectively. Different job-dispatching gives different empty-period
distribution. It turns out a more skew empty-period distribution leads
to more energy saving.

The intuition is that job-dispatching should try to make every server
empty as long as possible so that the on-off option, if explored,
can save abundant energy.
\begin{itemize}
\item Upon being assigned an empty period, a server only needs to \emph{independently}
make locally energy-optimal decision.
\end{itemize}
It is straightforward to verify that in the example, upon a job leaving
server S1 at time $0$, the procedure implicitly assigns an empty-period
of $T$ to S1, and turns S1 off if $\Delta<T$ and keeps it running
at idle state otherwise. Similarly, upon a job leaving S2 at time
$T_{2}^{c}$, S2 is turned off if $\Delta<\delta_{1}$ and stays idle
otherwise. Such comparisons and decisions can be done by individual
servers themselves.

\subsection{Offline Algorithm Achieving the Optimal Solution}

The \textbf{Optimal Solution} \textbf{Construction Procedure} determines
how many running servers to maintain at time $t$, i.e., $x^{*}(t)$,
to achieve the optimal server operation cost $P^{*}$. However, as
discussed in Section \ref{ssec:settings}, under the continuous-time
brick model, scheduling servers on/off according to $x^{*}(t)$ might
incur non-trivial job migration cost.

Exploiting the two observations made in the case-study at the end
of last subsection, we design a simple and decentralized off-line
algorithm that gives an optimal $x^{*}(t)$ and \emph{incurs no job
migration cost}.\\
\rule{1\columnwidth}{1pt}\\
\textbf{Decentralized Off-line Algorithm} \textbf{A0}:

\noindent \textbf{By a central job-dispatching entity}: it implements
a last-empty-server-first strategy. In particular, it maintains a
stack (i.e., a Last-In/First-Out queue) storing the IDs for all idle
or off servers. Before time $0$, the stack contains IDs for all the
servers that are not serving.
\begin{itemize}
\item Upon a job arrival: the entity pops a server ID from the top of the
stack, and assigns the job to the corresponding server (if the server
is off, the entity turns it on).
\item Upon a job departure: a server just turns idle, the entity pushes
the server ID into the stack.
\end{itemize}
\noindent\textbf{By each server:}
\begin{itemize}
\item Upon receiving a job: the server starts serving the job immediately.
\item Upon a job leaving this server and it becomes empty: let the current
time be $t_{1}$. The server searches for the earliest time $t_{2}\in(t_{1},t_{1}+\Delta]$
so that $a(t_{2})=a(t_{1})$. If no such $t_{2}$ exists, then the
server turns itself off. Otherwise, it stays idle.
\end{itemize}
\rule{1\columnwidth}{1pt}

We remark that in the algorithm, we use the same server to serve a
job during its entire sojourn time. Thus there is no job migration
cost. The following theorem justifies the optimality of the off-line
algorithm.
\begin{thm}
The proposed off-line algorithm \textbf{A0 }achieves the optimal server
operation cost of the problem \textbf{SCP}.\label{thm: offline}\end{thm}
\begin{IEEEproof}
Refer to Appendix \ref{apx:proof_5}.
\end{IEEEproof}
There are two important observations. First, the job-dispatching strategy
only depends on the past job arrivals and departures. Consequently,
the strategy assigns a job to the same server no matter it knows future
job arrival/departure or not; it also acts independently to servers'
off-or-idle decisions. Second, each individual server is actually
solving a classic ski-rental problem \cite{karlin1988competitive}
-- whether to {}``rent'', i.e., keep idle, or to {}``buy'', i.e.,
turn off now and on later, but with\emph{ }their {}``days-of-skiing''
(corresponding to servers' empty periods)\emph{ jointly determined
by the job-dispatching strategy}.

Next, we exploit these two observations to extend the off-line algorithm
\textbf{A0} to its online versions with performance guarantee.

\section{Online Dynamic Provisioning with or without Future Workload Information\label{sec:online}}

Inspired by our off-line algorithm, we construct online algorithms
by combining (i) the same last-empty-server-first job-dispatching
strategy as the one in algorithm \textbf{A0}, and (ii) an off-or-idle
decision module running on each server to \emph{solve an online ski-rental
problem}.

As discussed at the end of last section, the last-empty-server-first
job-dispatching strategy utilizes only past job arrival/departure
information. Consequently, as compared to the offline case, in the
online case it assigns the same set of jobs to the same server at
the same sequence of epochs. The following lemma rigorously confirms
this observation.
\begin{lem}
For the same $a\left(t\right),t\in\left[0,T\right]$, under the last-empty-server-first
job-dispatching strategy, each server will get the same job at the
same time and the job will leave the server at the same time for both
off-line and online situations. \label{lem:For-the-same}\end{lem}
\begin{IEEEproof}
Refer to Appendix \ref{apx:proof_6}.
\end{IEEEproof}
As a result, \emph{in the online case, each server still faces the
same set of off-or-idle problems} as compared to the off-line case.
This is the key to derive the competitive ratios of our to-be-presented
online algorithms.

Each server, not knowing the empty periods ahead of time, however,
needs to decide whether to stay idle or be off (and if so when) in
an online fashion. One natural approach is to adopt classic algorithms
for the online ski-rental problem.

\subsection{Dynamic Provisioning without Future Workload Information}

For the online ski-rental problem, the break-even algorithm in \cite{karlin1988competitive}
and the randomized algorithm in \cite{karlin1994competitive} have
competitive ratios $2$ and $e/\left(e-1\right)$, respectively. The
ratios have been proved to be optimal for deterministic and randomized
algorithms, respectively. Directly adopting these algorithms in the
off-or-idle decision module leads to two online solutions for the
problem \textbf{SCP} with competitive ratios $2$ and $e/\left(e-1\right)\approx1.58$.
These ratios improve the best known ratio $3$ achieved by the algorithm
in \cite{lin2011dynamic}.

The resulting solutions are decentralized and easy to implement: a
central entity runs the last-empty-server-first job-dispatching strategy,
and each server independently runs an online ski-rental algorithms.
For example, if the break-even algorithm is used, a server that just
becomes empty at time $t$ will stay idle for $\Delta$ amount of
time. If it receives no job during this period, it turns itself off.
Otherwise, it starts to serve the job immediately. As a special case
covered by Theorem \ref{thm:online}, it turns out this directly gives
a $2$-competitive dynamic provisioning solution.

\subsection{Dynamic Provisioning with Future Workload Information}

Classic online problem studies usually assume zero future information.
However, in our data center dynamic provisioning problem, one key
observation many existing solutions exploited is that the workload
expressed highly regular patterns. Thus the workload information in
a near prediction window may be accurately estimated by machine learning
or model fitting based on historical data \cite{chen2008energy,bod2009statistical}.
Can we exploit such future knowledge, if available, in designing online
algorithms? If so, how much gain can we get?

Let's elaborate through an example to explain why and how much future
knowledge can help. Suppose at any time $t$, the workload information
$a(t)$ in a prediction window $[t,t+\alpha\Delta]$ is available,
where $\alpha\in[0,1]$ is a constant. Consider a server running the
break-even algorithm just becomes empty at time $t_{1}$, and its
empty period happens to be just a bit longer than $\Delta$.

Following the standard break-even algorithm, the server waits for
$\Delta$ amount of time before turning itself off. According to the
setting, it receives a job right after $t_{1}+\Delta$ epoch, and
it has to power up to serve the job. This incurs a total cost of $2P\Delta$
as compared to the optimal one $P\Delta$, which is achieved by the
server staying idle all the way.

An alternative strategy that costs less is as follows. The server
stays idle for $\left(1-\alpha\right)\Delta$ amount of time, and
peeks into the prediction window $[t_{1}+\left(1-\alpha\right)\Delta,t_{1}+\Delta]$.
Due to the last-empty-server-first job-dispatching strategy, the server
can easy tell that it will receive a job if any $a(t)$ in the window
exceeds $a(t_{1})$, and no job otherwise. According to the setting,
the server sees itself receiving no job during $[t_{1}+\left(1-\alpha\right)\Delta,t_{1}+\Delta]$
and it turns itself off at time $t_{1}+\left(1-\alpha\right)\Delta$.
Later it turns itself on to serve the job right after $t_{1}+\Delta$.
Under this strategy, the overall cost is $\left(2-\alpha\right)P\Delta$
and is better than that of the break-even algorithm.

This simple example shows it is possible to modify classic online
algorithms to exploit future workload information to obtain better
performance. To this end, we propose new future-aware online ski-rental
algorithms and build new online solutions.

We model the availability of future workload information as follows.
For any $t$, the workload $a(t)$ for in the window $[t,t+\alpha\Delta]$
is known, where $\alpha\in[0,1]$ is a constant and $\alpha\Delta$
represents the size of the window.

We present both the modified break-even algorithm and the resulting
decentralized\emph{ }and\emph{ deterministic} online solution as follow.
The modified future-aware break-even algorithm is very simple and
is summarized as the part in the server's actions upon job departure.

\noindent\rule{1\columnwidth}{1pt}\\
\textbf{Future-Aware Online Algorithm A1:}

\noindent \textbf{By a central job-dispatching entity}: it implements
the last-empty-server-first job-dispatching strategy, i.e., the one
described in the off-line algorithm.

\noindent\textbf{By each server:}
\begin{itemize}
\item Upon receiving a job: the server starts serving the job immediately.
\item Upon a job leaving this server and it becomes empty: the server waits
for $\left(1-\alpha\right)\Delta$ amount of time,

\begin{itemize}
\item if it receives a job during the period, it starts serving the job
immediately;
\item otherwise, it looks into the prediction window of size $\alpha\Delta$.
It turns itself off, if it will receive no job during the window.
Otherwise, it stays idle.
\end{itemize}
\end{itemize}
\rule{1\columnwidth}{1pt}

In fact, as shown in Theorem \ref{thm:online} later in this section,
the algorithm \textbf{A1} has the best possible competitive ratio
for any deterministic algorithms under the last-empty-server-first
job-dispatching strategy. Thus, unless we change the job-dispatching
strategy, no deterministic algorithms can achieve better competitive
ratio than the algorithm \textbf{A1}.

Similarly, we present both the modified randomized algorithms for
solving online ski-rental problem and the resulting decentralized\emph{
}and\emph{ randomized} online solutions as follow. The modified future-aware
randomized algorithms are also summarized as the part in the server's
actions upon job departure. The first randomized algorithm \textbf{A2}
is a direct extension of the one in \cite{karlin1994competitive}
to make it future-aware. The algorithm \textbf{A3} is new and it has
the best possible competitive ratio for any randmonized algorithms
under the last-empty-server-first job-dispatching strategy.

\noindent\rule{1\columnwidth}{1pt}\\
\textbf{Future-Aware Online Algorithm A2:}

\noindent \textbf{By a central job-dispatching entity}: it implements
the last-empty-server-first job-dispatching strategy, i.e., the one
described in the off-line algorithm.

\noindent\textbf{By each server:}
\begin{itemize}
\item Upon receiving a job: the server starts serving the job immediately.
\item Upon a job leaving this server and it turns empty: the server waits
for $Z$ amount of time, where $Z$ is generated according to the
following probability density function
\[
f_{Z}(z)=\begin{cases}
\frac{e^{z/\left(1-\alpha\right)\Delta}}{\left(e-1\right)\left(1-\alpha\right)\Delta}, & \mbox{if }0\leq z\leq\left(1-\alpha\right)\Delta;\\
0, & \mbox{otherwise.}
\end{cases}
\]

\begin{itemize}
\item if it receives a job during the period, it starts serving the job
immediately;
\item otherwise, it looks into the prediction window of size $\alpha\Delta$.
It turns itself off, if it will receive no job during the window.
Otherwise, it stays idle.
\end{itemize}
\end{itemize}
\rule{1\columnwidth}{1pt}

\noindent\rule{1\columnwidth}{1pt}\\
\textbf{Future-Aware Online Algorithm A3:}

\noindent \textbf{By a central job-dispatching entity}: it implements
the last-empty-server-first job-dispatching strategy, i.e., the one
described in the off-line algorithm.

\noindent\textbf{By each server:}
\begin{itemize}
\item Upon receiving a job: the server starts serving the job immediately.
\item Upon a job leaving this server and it turns empty: the server waits
for $Z$ amount of time, where $Z$ is generated according to the
following probability distribution
\[
\begin{cases}
f_{Z}(z)=\begin{cases}
\frac{1-\frac{\alpha}{e-1+\alpha}}{\left(e-1\right)\vartriangle\left(1-\alpha\right)}e^{z/\left(1-\alpha\right)\Delta}, & \mbox{if }0<z\leq\left(1-\alpha\right)\Delta;\\
0, & \textrm{otherwise.}
\end{cases}\\
P\left(Z=0\right)=1-\frac{\alpha}{e-1+\alpha}
\end{cases}
\]

\begin{itemize}
\item if it receives a job during the period, it starts serving the job
immediately;
\item otherwise, it looks into the prediction window of size $\alpha\Delta$.
It turns itself off, if it will receive no job during the window.
Otherwise, it stays idle.
\end{itemize}
\end{itemize}
\rule{1\columnwidth}{1pt}

The three future-aware online algorithms inherit the nice properties
of the proposed off-line algorithm in the previous section. The same
server is used to serve a job during its entire sojourn time. Thus
there is no job migration cost. The algorithms are decentralized,
making them easy to implement and scale.

\begin{figure}
\centering\includegraphics[scale=0.2]{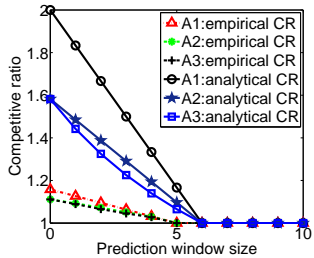}

\caption{Comparison of the worst-case competitive ratios (according to Theorem
\ref{thm:online}) and the empirical competitive ratios observed in
simulations using real-world traces. The critical window size $\Delta=6$
units of time. More simulation details are in Section \ref{sec:expr}.}
\label{fig:competitive_ratios}
\end{figure}

Observing no such future-aware online algorithms available in the
literature, we analyze their competitive ratios and present the results
as follows.
\begin{thm}
The deterministic online algorithm \textbf{A1} has a competitive ratio
of $2-\alpha$. The randomized online algorithm \textbf{A2} achieves
a competitive ratio of $\left(e-\alpha\right)/\left(e-1\right)$.
The randomized online algorithm \textbf{A3} achieves a competitive
ratio of $e/\left(e-1+\alpha\right)$. The competitive ratios of the
algorithms \textbf{A1} and are \textbf{A3} the best possible for deterministic
and randomized algorithms, respectively, under the last-empty-server-first
job-dispatching strategy. \label{thm:online}\end{thm}
\begin{IEEEproof}
Refer to Appendix \ref{apx:proof_6}.
\end{IEEEproof}
\textbf{Remarks}: (i) When $\alpha=1$, all three algorithms achieve
the optimal server operation cost. This matches the intuition that
servers only need to look $\Delta$ amount of time ahead to make optimal
off-or-idle decision upon job departures. This immediately gives a
fundamental insight that future workload information beyond the critical
interval $\Delta$ (corresponding to $\alpha=1$) will not improve
dynamic provisioning performance. (ii) The competitive ratios presented
in the above theorem is for the worst case. We have carried out simulations
using real-world traces and found the empirical ratios are much better,
as shown in Fig. \ref{fig:competitive_ratios}. (iii) To achieve better
competitive ratios, the theorem says that it is necessary to change
the job-dispatching strategy, since otherwise no deterministic or
randomized algorithms do better than the algorithms \textbf{A1} and
\textbf{A3}. (iv) Our analysis assumes the workload information in
the prediction window is accurate. We evaluate the two online algorithms
in simulations using real-world traces with prediction errors, and
observe they are fairly robust to the errors. More details are provided
in Section\textcolor{blue}{{} }\ref{sec:expr}.

\subsection{Adapting the Algorithms to Work with Discrete-Time Fluid Workload
Model \label{ssec:dis}}

Adapting our off-line and online algorithms to work with the discrete-time
fluid workload model involves two simple modifications. Recall in
the discrete-time fluid model, time is chopped into equal-length slots.
Jobs arriving in one slot get served in the same slot. Workload can
be split among running servers at arbitrary granularity like fluid.

For the job-dispatching entity in all the algorithms, at the end of
each slot when all servers are considered to be empty, it pushes all
the server IDs back into the stack (order doesn't matter). Then at
the beginning of each slot, it pops just-enough server IDs from the
stack in a Last-In/First-Out manner to satisfy the current workload.
In this way, the job-dispatching entity essentially packs the workload
to as few servers as possible, following the last-empty-server-first
strategy.

For individual servers, they start to serve upon receiving jobs, and
start to solve the off-line or online ski-rental problems upon all
its jobs leaving and it becomes empty.

It is not difficult to verify the modified algorithms still retain
their corresponding performance guarantees. Actually, we have following
corollary.
\begin{cor}
The modified deterministic and randomized online algorithms for discrete-time
fluid workload have competitive ratios of $2-\alpha$, $(e-\alpha)/(e-1)$,
and $e/\left(e-1+\alpha\right)$, respectively. \label{cormodified}\end{cor}
\begin{IEEEproof}
Refer to Appendix \ref{apx:proof_9}.
\end{IEEEproof}

\subsection{Comparison with the DELAYEDOFF Algorithm\label{ssec:comparison.with.DELAYEDOFF}}

It is somewhat surprising to find out our algorithms share similar
ingredients as the DELAYEDOFF algorithm in \cite{gandhi2010optimality},
since these are two independent efforts setting off to optimize different
objective functions (total energy consumption in our study v.s. Energy-Response
time Product (ERP) in \cite{gandhi2010optimality}).

The DELAYEDOFF algorithm contains two modules. The first one is a
job-dispatching module that assigns a newly arrived job to the most-recently-busy
idle server (i.e., the idle server who was most recently busy); servers
in off-state are not included. The second one is a delay-off module
running on each server that keeps the server idle for some pre-determined
amount of time, defined as $t_{wait}$, before turning it off. If
the server gets a job to service in this period, its idle time is
reset to $0$. The authors of \cite{gandhi2010optimality} show that
for any $t_{wait}$, if the job arrival process is Poisson, the DELAYEDOFF
algorithm minimizes the average ERP of a data center as the load (i.e.,
the ratio between the arrival rate and the average sojourn time) approaches
infinity.

Interestingly, if there are idle servers in system, DELAYEDOFF and
the algorithm \textbf{A1} will choose the same server to serve the
new job because the most-recently-busy server is indeed the last-empty
server in this case. If there are no idle servers, the algorithm \textbf{A1
}will still choose the last-empty server but DELAYEDOFF will randomly
select an off server to server the job. With this observation, the
DELAYEDOFF algorithm, under the setting $t_{wait}=\Delta$, can be
viewed as a variant of a special case of the algorithm \textbf{A1}
with zero future workload information available (i.e., $\alpha=0$).
It would be interesting to see whether the analytical insights used
in analyzing the DELAYEDOFF algorithm can be used to understand the
performance of the algorithm \textbf{A1} when the job arrival process
is Poisson.

Despite the similarity between the algorithm \textbf{A1} and the DELAYEDOFF
algorithm, it is not clear what is the competitive ratio of DELAYEDOFF.
Unlink our last-empty-server-first job-dispatching strategy, the most-recently-busy
idle server first strategy does not guarantee a server faces the same
set of ski-rental problems in the online case as compared to the off-line
case. Consequently, it is not clear how to relate the online cost
of the DELAYEDOFF algorithm to the offline optimal cost.

The two job-dispatching strategies differ more when the server waiting
time is random, e.g., in our algorithms \textbf{A2} and \textbf{A3},
where a later-empty server may turn itself off before an early-empty
server does; hence, the most-recently-busy (idle) server is usually
not the last-empty server. We compare the performance of algorithms
\textbf{A1}, \textbf{A2}, \textbf{A3}, and DELAYEDOFF in simulations
in Section \ref{sec:expr}.

\begin{center}
\begin{figure*}
\begin{centering}
\subfloat[MSR data trace for one week\label{fig:MSR}]{\includegraphics[width=0.25\textwidth]{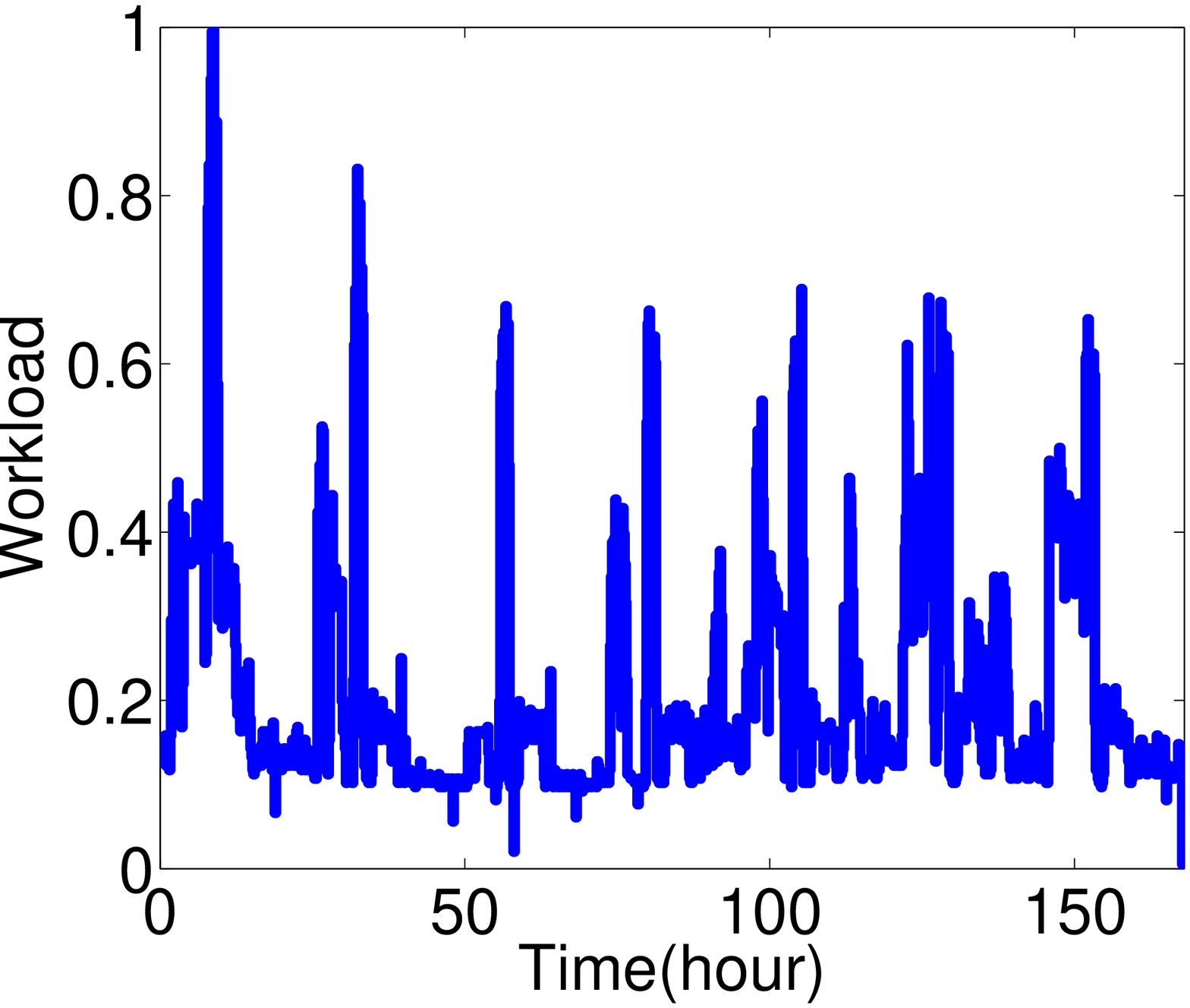}}\subfloat[Impact of future information\label{fig:Impact-of-future}]{\includegraphics[width=0.25\textwidth]{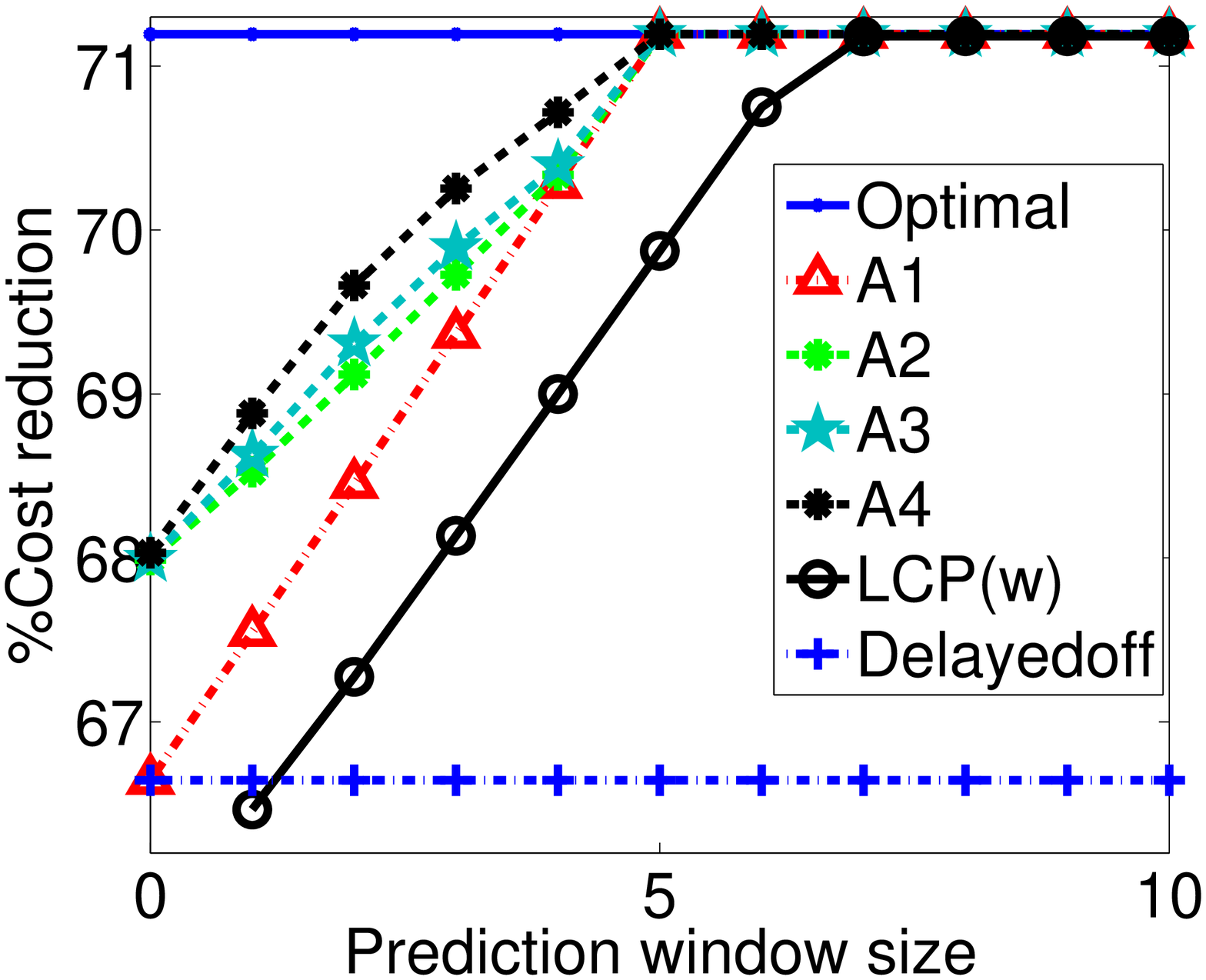}}\subfloat[Impact of prediction error\label{fig:Impact-of-prediction}]{\includegraphics[width=0.25\textwidth]{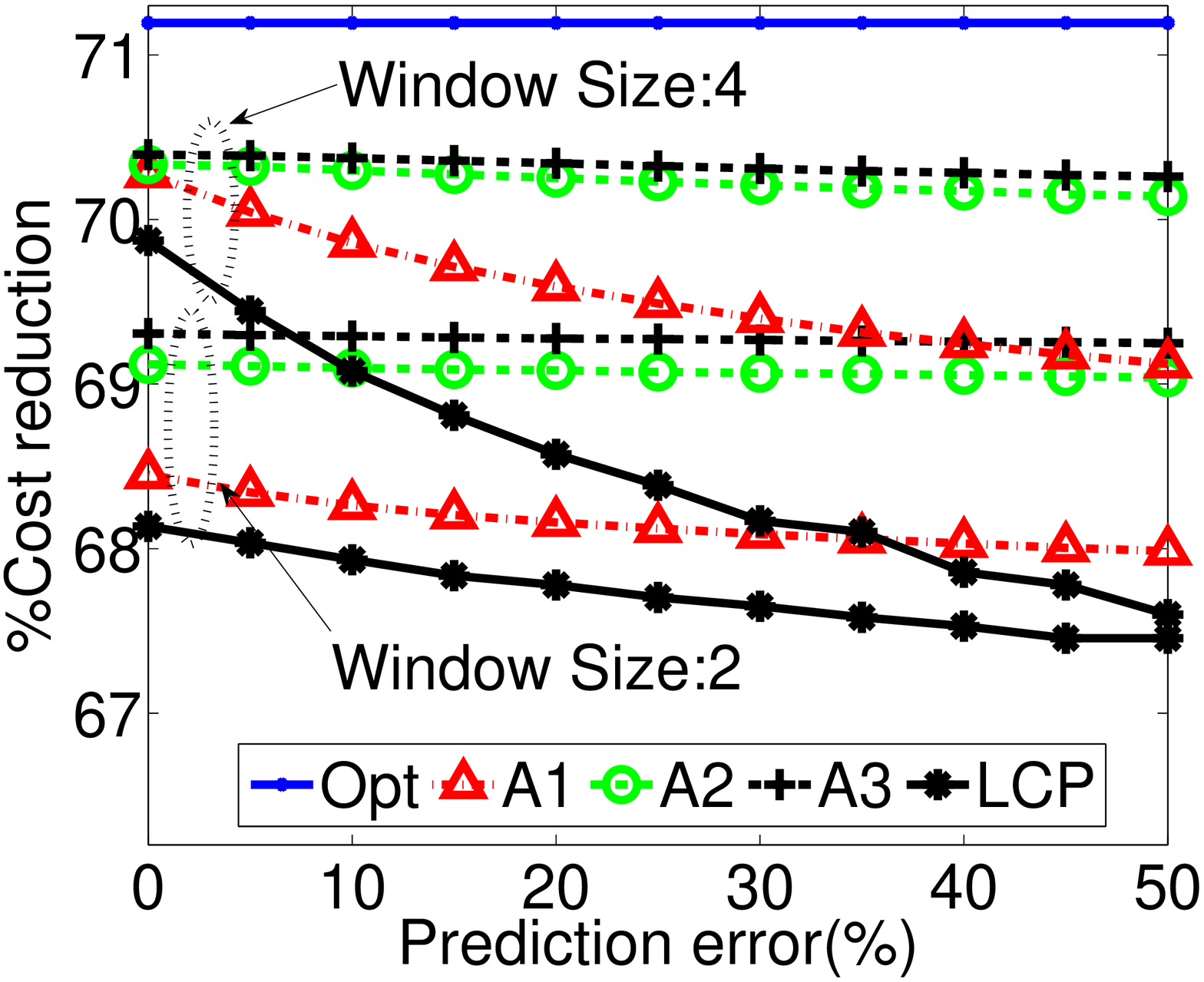}}\subfloat[Impact of PMR\label{fig:Impact-of-PMR}]{\includegraphics[width=0.25\textwidth]{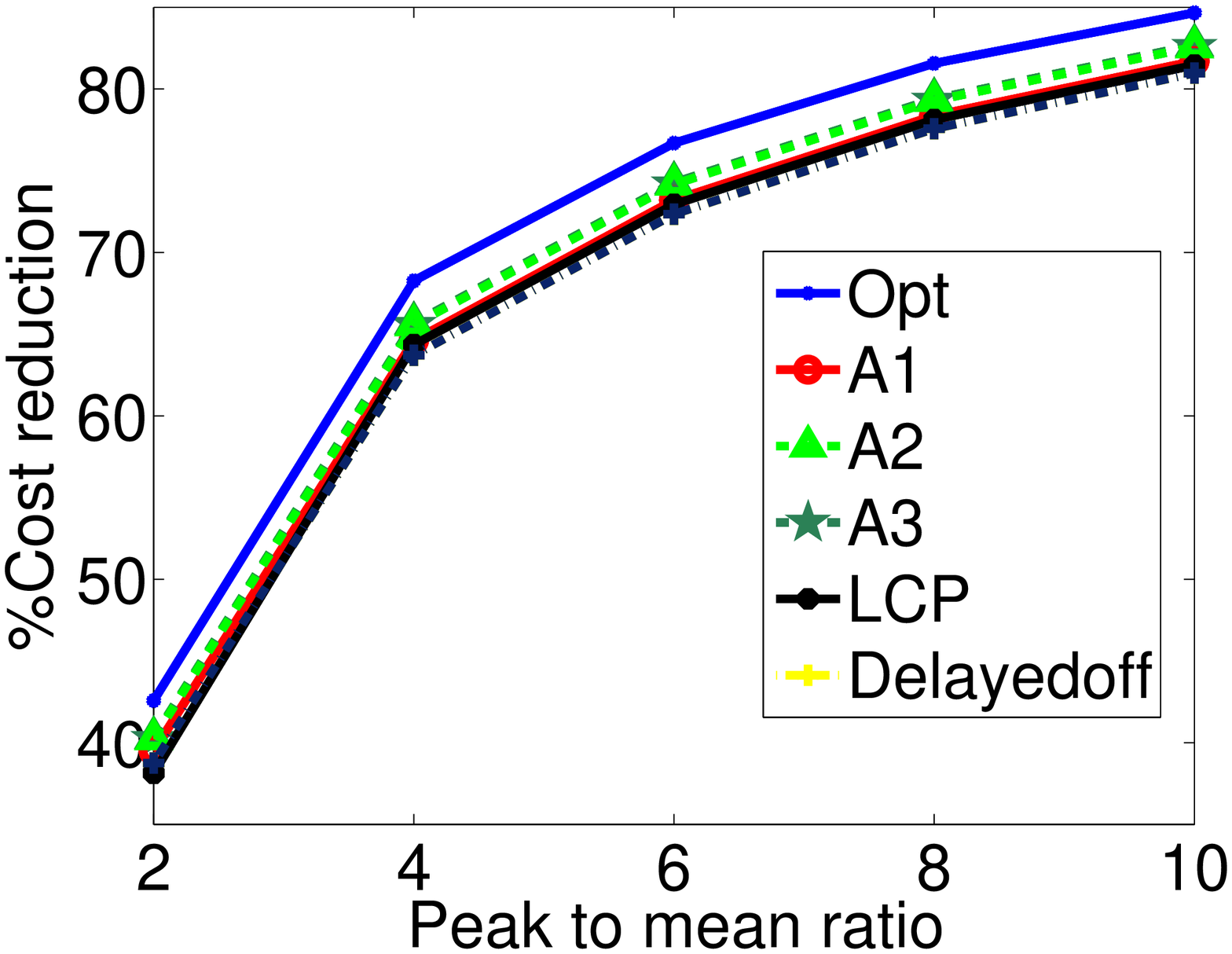}

}
\par\end{centering}

\centering{}\caption{Real-world workload trace and performance of the algorithms under
different situations.}
\end{figure*}

\par\end{center}

\section{Experiments \label{sec:expr}}

We implement the proposed off-line and online algorithms and carry
out simulations using real-world traces to evaluate their performance.
Our purposes are threefold. First, to evaluate the performance of
the algorithms using real-world traces. Second, to study the impacts
of workload prediction error and workload characteristic on the algorithms'
performance. Third, to compare our algorithms to two recently proposed
solutions LCP$(w$) in \cite{lin2011dynamic} and DELAYEDOFF in \cite{gandhi2010optimality}.

\subsection{Settings}

\textit{Workload trace}: The real-world traces we use in experiments
are a set of I/O traces taken from 6 RAID volumes at MSR Cambridge
\cite{narayanan2008write}. The traced period was one week between
February 22 to 29, 2007. We estimate the average number of jobs over
disjoint 10 minute intervals. The data trace has a peak-to-mean ratio
(PMR) of 4.63. The jobs are {}``request-response'' type and thus
the workload is better described by a discrete-time fluid model, with
the slot length being 10 minutes and the load in each slot being the
average number of jobs.

As discussed in Section \ref{ssec:dis}, the proposed off-line and
online algorithms also work with the discrete-time fluid workload
model after simple modification. In the experiments, we run the modified
algorithms using the above real-world traces.

\textit{Cost benchmark}: Current data centers usually do not use dynamic
provisioning. The cost incurred by static provisioning is usually
considered as benchmark to evaluate new algorithms \cite{lin2011dynamic,krioukov2011napsac}.
Static provisioning runs a constant number of servers to serve the
workload. In order to satisfy the time-varying demand during a period,
data centers usually overly provision and keep more running servers
than what is needed to satisfy the peak load. In our experiment, we
assume that the data center has the complete workload information
ahead of time and provisions exactly to satisfy the peak load. Using
such benchmark gives us a conservative estimate of the cost saving
from our algorithms.

\textit{Sever operation cost:} The server operation cost is determined
by unit-time energy cost $P$ and on-off costs $\beta_{on}$ and $\beta_{off}$.
In the experiment, we assume that a server consumes one unit energy
for per unit time, i.e., $P=1$. We set $\beta_{off}+\beta_{on}=6$,
i.e., the cost of turning a server off and on once is equal to that
of running it for six units of time \cite{lin2011dynamic}. Under
this setting, the critical interval is $\Delta=\left(\beta_{off}+\beta_{on}\right)/P=6$
units of time.

\subsection{\textit{\emph{Performance of the Proposed Online Algorithms}}}

We have characterized in Theorem \ref{thm:online} the competitive
ratios of our proposed online algorithms as the prediction window
size, i.e., $\alpha\Delta$, increases. The resulting competitive
ratios, i.e., $2-\alpha$, $\left(e-\alpha\right)/\left(e-1\right)$
and $e/\left(e-1+\alpha\right)$, already appealing, are for the worst-case
scenarios. In practice, the actual performance can be even better.

In our first experiment, we study the performance of our online algorithms
using real-world traces. The results are shown in Fig. \ref{fig:Impact-of-future}.
The cost reduction curves are obtained by comparing the power cost
incurred by the off-line algorithm, the three online algorithms, the
LCP$\left(w\right)$ algorithm \cite{lin2011dynamic} and the DELAYEDOFF
algorithm \cite{gandhi2010optimality} to the cost benchmark. The
vertical axis indicates the cost reduction and the horizontal axis
indicates the size of prediction window varying from 0 to 10 units
of time.

As seen, for this set of workload, both our three online algorithms,
LCP$\left(w\right)$ and DELAYEDOFF achieve substantial cost reduction
as compared to the benchmark. In particular, the cost reductions of
our three online algorithms are beyond $66\%$ even when no future
workload information is available; while LCP$\left(w\right)$ has
to have (or estimate) one unit time of future workload to execute,
and thus it starts to perform when the prediction window size is one.
The cost reductions of our three online algorithms grow linearly as
the prediction window increases, and reaching optimal when the prediction
window size reaches $\Delta$. These observations match what Theorem
\ref{thm:online} predicts. Meanwhile, LCP$\left(w\right)$ has not
yet reach the optimal performance when the prediction window size
reaches the critical value $\Delta$.  DELAYEDOFF has the same performance
for all prediction window sizes since it does not exploit future workload
information.

As seen in Fig. \ref{fig:Impact-of-future}, in the simulation, our
three algorithms can achieve the optimal power consumption when the
size of prediction window is $5$, one unit smaller than the theoretically-computed
one $\Delta=6$. At first glance, the results seem not aligned with
what the analysis suggests. But a careful investigation reveals that
there is no mis-alignment between analysis and simulation. Because
jobs are assigned to servers at the beginning of each slots in discrete-time
fluid model, knowing the workload from current time to the beginning
of the 5th look-ahead future slot is equivalent to knowing the workload
of a duration of 6 slots. Hence, the anaysis indeed suggests Algorithms
\textbf{A1-A3} can achieve optimal power consumption when the size
of prediction window is $5$, as observed in Fig. \ref{fig:Impact-of-future}.

\subsection{\textit{\emph{Impact of Prediction Error}}\emph{ }}

Previous experiments show that both our algorithms and LCP$\left(w\right)$
have better performance if accurate future workload is available.
However, there are always prediction errors in practice. Therefore,
it is important to evaluate the performance of the algorithms in the
present of prediction error.

To achieve this goal, we evaluate our online algorithms with prediction
window size of 2 and 4 units of time. Zero-mean Gaussian prediction
error is added to each unit-time workload in the prediction window,
with its standard deviation grows from $0$ to $50\%$ of the corresponding
actual workload. In practice, prediction error tends to be small \cite{kusic2009power};
thus we are essentially stress-testing the algorithms.

We average 100 runs for each algorithm and show the results in Fig.
\ref{fig:Impact-of-prediction}, where the vertical axis represents
the cost reduction as compared to the benchmark.

On one hand, we observe all algorithms are fairly robust to prediction
errors. On the other hand, all algorithms achieve better performance
with prediction window size 4 than size 2. This indicates more future
workload information, even inaccurate, is still useful in boosting
the performance.

\subsection{\textit{\emph{Impact of Peak-to-Mean Ratio (PMR)}}}

Intuitively, comparing to static provisioning, dynamic provisioning
can save more power when the data center trace has large PMR. Our
experiments confirm this intuition which is also observed in other
works \cite{lin2011dynamic,krioukov2011napsac}. Similar to \cite{lin2011dynamic},
we generate the workload from the MSR traces by scaling $a\left(t\right)$
as $\overline{a\left(t\right)}=Ka^{\gamma}\left(t\right)$, and adjusting
$\gamma$ and $K$ to keep the mean constant. We run the off-line
algorithm, the three online algorithms, LCP$\left(w\right)$ and DELAYEDOFF
using workloads with different PMRs ranging from 2 to 10, with prediction
window size of one unit time. The results are shown in Fig. \ref{fig:Impact-of-PMR}.

As seen, energy saving increases form about $40\%$ at PRM=2, which
is common in large data centers, to large values for the higher PMRs
that is common in small to medium sized data centers. Similar results
are observed for different prediction window sizes.

\section{Concluding Remarks}

\label{sec:conclusion}

Dynamic provisioning is an effective technique in reducing server
energy consumption in data centers, by turning off unnecessary servers
to save energy. In this paper, we design online dynamic provisioning
algorithms with zero or partial future workload information available.

We reveal an elegant {}``divide-and-conquer'' structure of the off-line
dynamic provisioning problem, under the cost model that a running
server consumes a fixed amount energy per unit time. Exploiting such
structure, we show its optimal solution can be achieved by the data
center adopting a simple last-empty-server-first job-dispatching strategy
and each server independently solving a classic ski-rental problem.

We build upon this architectural insight to design two new decentralized
online algorithms. One is a deterministic algorithm with competitive
ratio $2-\alpha$, where $0\leq\alpha\leq1$ is the fraction of a
critical window in which future workload information is available.
The size of the critical window is determined by the wear-and-tear
cost and the unit-time energy cost of running a single server. The
other two are randomized algorithms with competitive ratios $\left(e-\alpha\right)/\left(e-1\right)\approx1.58-\alpha/\left(e-1\right)$
and $e/\left(e-1+\alpha\right)$, respectively. $2-\alpha$ and $e/\left(e-1+\alpha\right)$
are the best competitive ratios for deterministic and randomized online
algorithms under our last-empty-server-first job-dispatching strategy.
Our results also lead to a fundamental observation that under the
cost model that a running server consumes a fixed amount energy per
unit time, future workload information beyond the critical window
will not improve the dynamic provisioning performance.

Our algorithms are simple and easy to implement. Simulations using
real-world traces show that our algorithms can achieve close-to-optimal
energy-saving performance, and are robust to future-workload prediction
errors.

Our results, together with the $3$-competitive algorithm recently
proposed by Lin \emph{et al.} \cite{lin2011dynamic}, suggest that
it is possible to reduce server energy consumption significantly with
zero or only partial future workload information.

An interesting and important future direction is to explore what is
the best possible competitive ratio any algorithms can achieve with
zero or partial future workload information. Insights along this line
provides useful understanding on the benefit of knowing future workload
in dynamic provisioning.

\section*{Acknowledgements}

We thank Minghong Lin and Lachlan Andrew for sharing the code of their
LCP algorithm, and Eno Thereska for sharing the MSR Cambridge data
center traces.

\appendix

\subsection{Proof of Proposition \textmd{\normalsize \ref{prop:property} \label{apx:proof_1}}}
\begin{IEEEproof}
The proof that critical segment $\left[T_{i}^{c},T_{i+1}^{c}\right]$
must belong to one of the four types described in proposition \ref{prop:property}
is based on two cases.

Case 1: $T_{i}^{c}$ is job-arrival epoch.

In this case, according to our \textbf{Critical Segment Construction
Procedure}, $T_{i+1}^{c}$ is the first departure epoch $\tau$ after
$T_{i}^{c}$. Then workload in $\left[T_{i}^{c},T_{i+1}^{c}\right]$
is non-decreasing, which means $\left[T_{i}^{c},T_{i+1}^{c}\right]$
is Type-I critical segment.

Case 2: $T_{i}^{c}$ is job-departure epoch.

In this case, we have two sub-cases. First, if we can find the first
arrival epoch $\tau$ after $T_{i}^{c}$ so that $a\left(\tau\right)=a\left(T_{i}^{c}\right)$,
according to \textbf{Critical Segment Construction Procedure}, we
let $T_{i+1}^{c}=\tau$. If $a\left(t\right)=a\left(T_{i}^{c}\right)-1,\forall t\in\left(T_{i}^{c},T_{i+1}^{c}\right)$,
$\left[T_{i}^{c},T_{i+1}^{c}\right]$ is Type-III critical segment.
Otherwise, $\left[T_{i}^{c},T_{i+1}^{c}\right]$ is Type-IV critical
segment, $a\left(T_{i+1}^{c}\right)=a\left(T_{i}^{c}\right)$, $a\left(t\right)\leq a\left(T_{i}^{c}\right)-1$
and not always identical, $\forall t\in\left(T_{i}^{c},T_{i+1}^{c}\right)$.
Second, if no such $\tau$ exists, then we let $T_{i+1}^{c}$ to be
the next job departure epoch, then $\left[T_{i}^{c},T_{i+1}^{c}\right]$
is Type-II critical segment. $a\left(t\right)$ in this segment is
step-decreasing, which means $a\left(t\right)=a\left(T_{i}^{c}\right)-1,\forall t\in\left(T_{i}^{c},T_{i+1}^{c}\right]$
and $a\left(t\right)\leq a\left(T_{i+1}^{c}\right)-1,\forall t\in\left(T_{i+1}^{c},T\right]$.

The above two cases cover all the possible situations of critical
segment $\left[T_{i}^{c},T_{i+1}^{c}\right]$. And we proved that
$\left[T_{i}^{c},T_{i+1}^{c}\right]$ must belong to one of the four
types for both cases. Hence, we proved proposition \ref{prop:property}. \end{IEEEproof}

\subsection{Proof of Lemma \ref{lem:lemma 2}\label{apx:proof_2}}
\begin{IEEEproof}
Because at $t=T_{1}^{c}=0,$ we have $x^{*}\left(0\right)=a\left(0\right)$,
which means $,x^{*}\left(t\right)$ meets $a\left(t\right)$ at the
first critical time. We will use induction to prove Lemma \ref{lem:lemma 2}
is true for all the rest critical times. As a matter of fact, given
$x^{*}\left(T_{i}^{c}\right)=a\left(T_{i}^{c}\right)$, we claim that
$x^{*}\left(T_{i+1}^{c}\right)=a\left(T_{i+1}^{c}\right)$. We divide
the situation in two cases and in each case we will prove $x^{*}\left(T_{i+1}^{c}\right)=a\left(T_{i+1}^{c}\right)$
by adopting proof-by-contradiction.

Case 1: When $\left[T_{i}^{c},T_{i+1}^{c}\right]$ is Type-I, Type-III
or Type-IV critical segment, which means we must have $a\left(T_{i}^{c}\right)\leq a\left(T_{i+1}^{c}\right)$.

If $x^{*}\left(T_{i+1}^{c}\right)>a\left(T_{i+1}^{c}\right)$, then
we can find a time $\tau\in\left[T_{i}^{c},T_{i+1}^{c}\right)$ such
that $x^{*}\left(\tau\right)=a\left(T_{i+1}^{c}\right)$ and $x^{*}\left(t\right)>a\left(T_{i+1}^{c}\right),\forall t\in\left(\tau,T_{i+1}^{c}\right]$.
Define $\overline{x\left(t\right)}$ as follows: $\overline{x\left(t\right)}=x^{*}\left(t\right),\forall t\in\left[0,\tau\right]\cup\left(T_{i+1}^{c},T\right]$
and $\overline{x\left(t\right)}=a\left(T_{1}^{c}\right),\forall t\in\left(\tau,T_{i+1}^{c}\right]$.
It is clear that $\overline{x\left(t\right)}$ satisfy the constraints
of \eqref{eq: obj}. Moreover, $x^{*}\left(t\right)$ will cause more
power consumption than $\overline{x\left(t\right)}$ because $x^{*}\left(t\right)$
will consume more power to run extra servers during $\left(\tau,T_{i+1}^{c}\right]$
and both have the same power consumption for the rest of time. It
is a contradiction that $x^{*}\left(t\right)$ is an optimal solution
of \eqref{eq: obj}. Therefore, $x^{*}\left(T_{i+1}^{c}\right)=a\left(T_{i+1}^{c}\right)$.

Case 2: When $\left[T_{i}^{c},T_{i+1}^{c}\right]$ is Type-II\textit{
}critical segment, which means we must have $a\left(T_{i}^{c}\right)>a\left(T_{i+1}^{c}\right)\geq a\left(T\right)$.

If $x^{*}\left(T_{i+1}^{c}\right)>a\left(T_{i+1}^{c}\right)$, because
$x^{*}\left(T\right)=a\left(T\right)\leq a\left(T_{i+1}^{c}\right)$,
then we can find a time $T_{i+1}^{c}<\tau\leq T$ such that $x^{*}\left(\tau\right)=a\left(T_{i+1}^{c}\right)$
and $x^{*}\left(t\right)>a\left(T_{i+1}^{c}\right),\forall t\in\left[T_{i+1}^{c},\tau\right)$.
Define $\overline{x\left(t\right)}$ as follows: $\overline{x\left(t\right)}=x^{*}\left(t\right),\forall t\in\left[0,T_{i+1}^{c}\right)\cup\left[\tau,T\right]$
and $\overline{x\left(t\right)}=a\left(T_{i+1}^{c}\right),\forall t\in\left[T_{i+1}^{c},\tau\right)$.
It is clear that $\overline{x\left(t\right)}$ satisfy the constraint
of \eqref{eq: obj} due to property \ref{enu:property 2} of Type-II\textit{
}critical segments. Moreover, $x^{*}\left(t\right)$ will cost more
power consumption than $\overline{x\left(t\right)}$ because $x^{*}\left(t\right)$
will consume more power to run extra servers during $\left[T_{i+1}^{c},\tau\right)$
and both have the same power consumption for the rest of time. It
is a contradiction that $x^{*}\left(t\right)$ is an optimal solution
of \eqref{eq: obj}. Therefore, $x^{*}\left(T_{i+1}^{c}\right)=a\left(T_{i+1}^{c}\right)$.

Above two cases cover all the possibility of critical segment $\left[T_{i}^{c},T_{i+1}^{c}\right]$
and we proved that $x^{*}\left(T_{i+1}^{c}\right)=a\left(T_{i+1}^{c}\right)$
in both two cases. Therefore, we proved Lemma \ref{lem:lemma 2}.\end{IEEEproof}

\subsection{Proof of Lemma \ref{prop: lower bound}\label{apx:proof_3}}
\begin{IEEEproof}
Let $P_{i}^{x^{*}}$ denote power consumption in critical segment
$\left[T_{i}^{c},T_{i+1}^{c}\right]$ if we let $x\left(t\right)=x^{*}\left(t\right),\forall t\in\left[T_{i}^{c},T_{i+1}^{c}\right]$.
According to the Lemma \ref{lem:lemma 2}, we have $x^{*}\left(T_{i}^{c}\right)=a\left(T_{i}^{c}\right)$
and $x^{*}\left(T_{i+1}^{c}\right)=a\left(T_{i+1}^{c}\right)$. Therefore,
$x^{*}\left(t\right),\forall t\in\left[T_{i}^{c},T_{i+1}^{c}\right]$
is a solution to optimization problem \eqref{eq:sub}. Thus, we have
$P_{i}^{x^{*}}\geq P_{i}^{*}$ and $P^{*}=\underset{i=1}{\overset{M-1}{\sum}}P_{i}^{x^{*}}\geq\sum P_{i}^{*}$.
This proves Lemma \ref{prop: lower bound}.\end{IEEEproof}

\subsection{Proof of Theorem \ref{Thm:opt_sol_pro_is_opt} \label{apx:proof_4}}

Before proving theorem \ref{Thm:opt_sol_pro_is_opt} , we first prove
following Lemma.

Define $\mathcal{P}\left(A,B,T_{s},T_{e}\right)$ as the following
optimization problem. $\left[T_{s},T_{e}\right]$ satisfy $a\left(T_{s}\right)=a\left(T_{e}\right)$,
$a\left(t\right)<a\left(T_{s}\right),\forall t\in\left(T_{s},T_{e}\right)$
and $\left(T_{e}-T_{s}\right)>\vartriangle$. $A,B$ are constants
which are greater than or equal to $a\left(T_{s}\right)$.

\begin{eqnarray}
 & \textrm{min} & P\varint_{T_{s}}^{T_{e}}x\left(t\right)dt+P_{on}\left(T_{s},T_{e}\right)+P_{off}\left(T_{s},T_{e}\right)\label{eq:ness}\\
 & \textrm{s.t}. & x(t)\geq a(t),\forall t\in\left[T_{s},T_{e}\right],\\
 &  & x(T_{s})=A,x(T_{e})=B,\\
 & \mbox{var} & x(t)\in\mathbb{Z}^{+},t\in\left[T_{s},T_{e}\right].
\end{eqnarray}

\begin{lem}
The necessary condition for $x\left(t\right)$ to achieve optimal
power consumption of $\mathcal{P}\left(A,B,T_{s},T_{e}\right)$ is
that $x\left(t\right)\leq a\left(T_{s}\right)-1,\forall t\in\left(T_{s},T_{e}\right)$
.\label{lem:necessary}\end{lem}
\begin{IEEEproof}
Let $x_{i}\left(t\right)$ be any optimal solution to above optimization
problem $\mathcal{P}\left(A,B,T_{s},T_{e}\right)$ and $x_{i}\left(t\right)$
does not satisfy $x_{i}\left(t\right)\leq a\left(T_{s}\right)-1,\forall t\in\left(T_{s},T_{e}\right)$.
In order to prove the necessary condition, we divide $x_{i}\left(t\right)$
into four cases.

$\left(a\right)$ $x_{i}\left(t\right)\geq a\left(T_{s}\right),\forall t\in\left(T_{s},T_{e}\right)$.

In this case, let $\overline{x}\left(t\right)=a\left(T_{s}\right)-1,\forall t\in\left(T_{s},T_{e}\right)$,
then $x_{i}\left(t\right)$ will consume at least $\left(T_{e}-T_{s}\right)P$
more power to run extra servers than $\overline{x}\left(t\right)$
during $\left(T_{s},T_{e}\right)$. On the other hand, $\overline{x}\left(t\right)$
causes at most $\beta_{on}+\beta_{off}$ more wear-and-tear cost than
$x_{i}\left(t\right)$. Because $\left(T_{e}-T_{s}\right)>\vartriangle$,
$x_{i}\left(t\right)$ actually cost more power than $\overline{x}\left(t\right)$,
which is a contradiction with that $x_{i}\left(t\right)$ is an optimal
solution.

$\left(b\right)$$\exists\tau\in\left(T_{s},T_{e}\right)$ such that
$x_{i}\left(\tau\right)=a\left(T_{s}\right)-1,x_{i}\left(t\right)>a\left(T_{s}\right)-1,\forall t\in\left(T_{s},\tau\right)$.

In this case, let $\overline{x}\left(t\right)=a\left(T_{s}\right)-1,\forall t\in\left(T_{s},\tau\right)$
and $\overline{x}\left(t\right)=x_{i}\left(t\right),\forall t\in\left[\tau,T_{e}\right)$.
then it is clear that $x_{i}\left(t\right)$ consume more power than
$\overline{x}\left(t\right)$, which is a contradiction with that
$x_{i}\left(t\right)$ is an optimal solution.

$\left(c\right)$ $\exists\tau\in\left(T_{s},T_{e}\right)$ such that
$x_{i}\left(\tau\right)=a\left(T_{s}\right)-1,x_{i}\left(t\right)>a\left(T_{s}\right)-1,\forall t\in\left(\tau,T_{e}\right)$

In this case, let $\overline{x}\left(t\right)=a\left(T_{s}\right)-1,\forall t\in\left(\tau,T_{e}\right)$
and $\overline{x}\left(t\right)=x_{i}\left(t\right),\forall t\in\left(T_{s},\tau\right]$.
then it is clear that $x_{i}\left(t\right)$ consume more power than
$\overline{x}\left(t\right)$, which is a contradiction with that
$x_{i}\left(t\right)$ is an optimal solution.

$\left(d\right)$ $x_{i}\left(t\right)$ dose not satisfy above three
cases.

If $x_{i}\left(t\right)$ does not satisfy case $\left(a\right)\left(b\right)\left(c\right)$,
then there must exist time $\tau_{1}$ and $\tau_{2}$ in $\left(T_{s},T_{e}\right)$
such that $x_{i}\left(\tau_{1}\right)=x_{i}\left(\tau_{2}\right)=a\left(T_{s}\right)-1$
and $x_{i}\left(t\right)>a\left(T_{s}\right)-1,\forall t\in\left(\tau_{1},\tau_{2}\right)$.
Let $\overline{x}\left(t\right)=x_{i}\left(t\right),\forall t\in\left(T_{s},\tau_{1}\right]\cup\left[\tau_{2},T_{e}\right)$
and $\overline{x}\left(t\right)=a\left(T_{s}\right)-1,\forall t\in\left(\tau_{1},\tau_{2}\right)$.
$\overline{x}\left(t\right)$ satisfies all the constraints of \eqref{eq:ness}.
It is also easy to verify that $x_{i}\left(t\right)$ consume more
power than $x\left(t\right)$, which is a contradiction with that
$x_{i}\left(t\right)$ is an optimal solution.

The above four cases cover all possible situation of $x_{i}\left(t\right)$.
Therefore, we proved that the necessary condition for $x\left(t\right)$
to be an optimal solution to \eqref{eq:ness} is that $x\left(t\right)\leq a\left(T_{s}\right)-1,\forall t\in\left(T_{s},T_{e}\right)$.
\end{IEEEproof}
Now we are going to prove theorem \ref{Thm:opt_sol_pro_is_opt}.
\begin{IEEEproof}
Let $\overline{x_{i}}^{*}\left(t\right)$ denote the number of running
server constructed by \textbf{Optimal Solution} \textbf{Construction
Procedure} in critical segment $\left[T_{i}^{c},T_{i+1}^{c}\right]$.
We will prove that $\overline{x_{i}}^{*}\left(t\right)$ is an optimal
solution of \eqref{eq:sub}. The proof is based on the type of critical
segment $\left[T_{i}^{c},T_{i+1}^{c}\right]$.

For critical segments of Type-I and Type-II, we claim that $\overline{x}_{i}^{*}\left(t\right)=a\left(t\right),\forall t\in\left(T_{i}^{c},T_{i+1}^{c}\right)$
can achieve $P_{i}^{*}$. Let $x_{i}\left(t\right)$ be any solution
to \eqref{eq:sub} and $x_{i}\left(t\right)$ is not always equal
to $a\left(t\right)$ during $\left(T_{i}^{c},T_{i+1}^{c}\right)$.
Because $a\left(t\right)$ is either non-decreasing or step-decreasing
in Type-I and Type-II critical segments, we can find periods $\left(t_{1},t_{2}\right)$
in $\left(T_{i}^{c},T_{i+1}^{c}\right)$ such that $a\left(t_{1}\right)=x_{i}\left(t_{1}\right)$,
$a\left(t_{2}\right)=x_{i}\left(t_{2}\right)$ and $x_{i}\left(t\right)>a\left(t\right),\forall t\in\left(t_{1},t_{2}\right)$.
One example of such period is $\left(t_{3},t_{4}\right)$ in Fig.
\ref{fig:example1}. It is clear that $x_{i}\left(t\right)$ cost
more power than $\overline{x_{i}}^{*}\left(t\right)$ in each such
period and both have the same power consumption in the rest of time
during $\left(T_{i}^{c},T_{i+1}^{c}\right)$. Therefore, $\overline{x_{i}}^{*}\left(t\right)$
is an optimal solution to \eqref{eq:sub} and can achieve optimal
power consumption $P_{i}^{*}$.

\begin{figure}
\begin{centering}
\includegraphics[width=8cm]{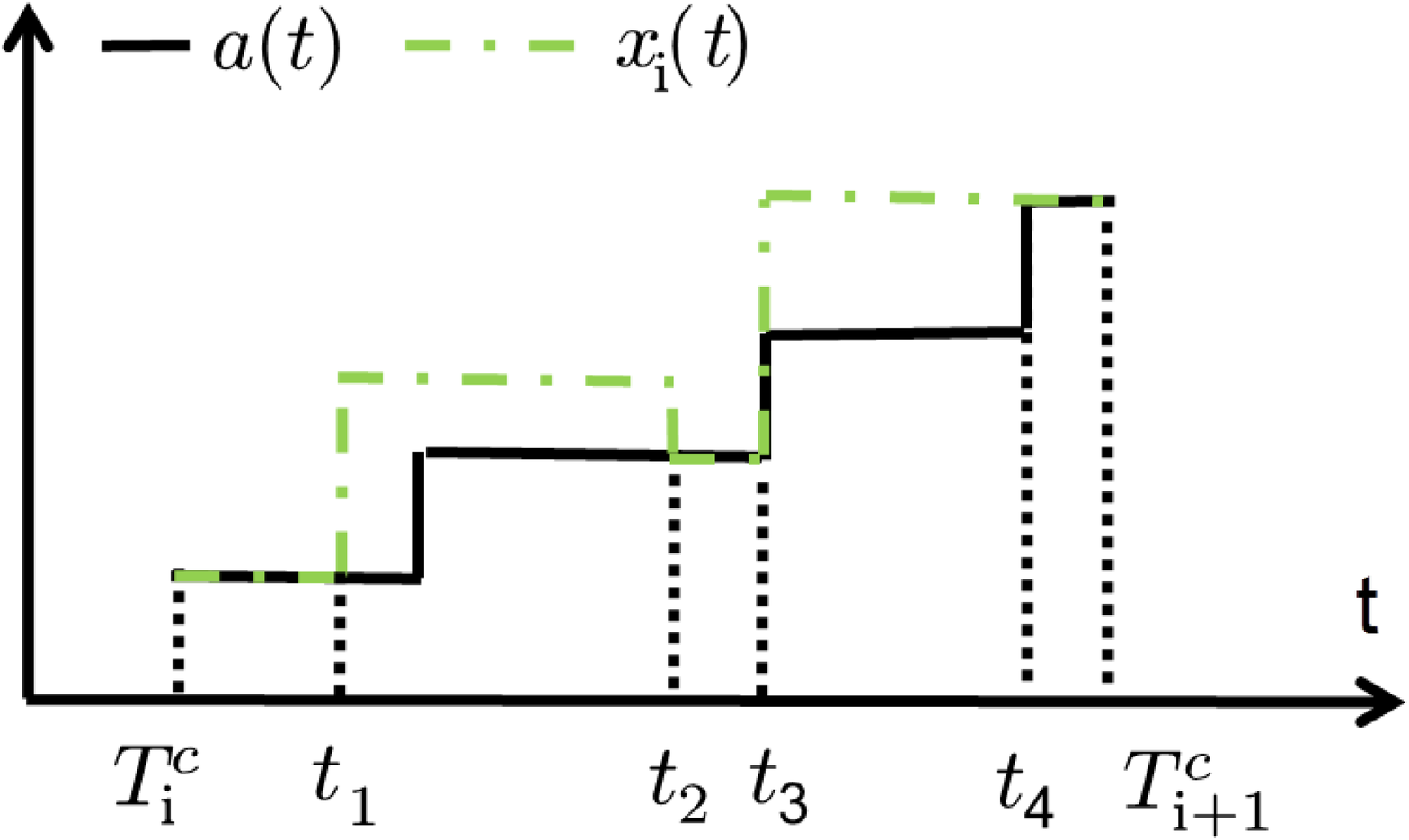}
\par\end{centering}

\caption{An example of solution $x_{i}\left(t\right)$ to \eqref{eq:sub} in
Type-I critical segment. $x_{i}\left(t\right)$ is greater than $a\left(t\right)$
in $\left(t_{1},t_{2}\right)$ and $\left(t_{3},t_{4}\right)$.\label{fig:example1}}
\end{figure}

It is clear that $\bar{x_{i}}^{*}\left(t\right)\leq a\left(T_{i}^{c}\right),\forall t\in\left(T_{i}^{c},T_{i+1}^{c}\right)$
for Type-III segment according to our \textbf{Optimal Solution} \textbf{Construction
Procedure}. We divide the proof of theorem \ref{Thm:opt_sol_pro_is_opt}
for Type-III critical segment in two cases.

Case 1: $\vartriangle\geq\left(T_{i+1}^{c}-T_{i}^{c}\right)$.

In this case, we claim that $\bar{x_{i}}^{*}\left(t\right)=a\left(T_{i}^{c}\right),\forall t\in\left(T_{i}^{c},T_{i+1}^{c}\right)$
can achieve $P_{i}^{*}$. In fact, let $x_{i}\left(t\right)$ be any
solution to \eqref{eq:sub} and $x_{i}\left(t\right)$ is not always
equal to $a\left(T_{i}^{c}\right)$ during $\left(T_{i}^{c},T_{i+1}^{c}\right)$.
We will prove that $\bar{x_{i}}^{*}\left(t\right)$ does not cost
more power consumption than $x_{i}\left(t\right)$ in $\left(T_{i}^{c},T_{i+1}^{c}\right)$.

Since $x_{i}\left(t\right)$ is not always equal to $a\left(T_{i}^{c}\right)$
during $\left(T_{i}^{c},T_{i+1}^{c}\right)$, we can find period $\left(t_{1},t_{2}\right)\subseteq\left(T_{i}^{c},T_{i+1}^{c}\right)$
such that $a\left(T_{i}^{c}\right)=x_{i}\left(t_{1}\right)$, $a\left(T_{i}^{c}\right)=x_{i}\left(t_{2}\right)$
and $x_{i}\left(t\right)\neq a\left(T_{i}^{c}\right),\forall t\in\left(t_{1},t_{2}\right)$.
One example of such period is $\left(t_{1},t_{2}\right)$ in Fig.
\ref{fig:example2}.

We will compare the power consumed by $x_{i}\left(t\right)$ and $\bar{x_{i}}^{*}\left(t\right)$
in $\left(t_{1},t_{2}\right)$ based on two situations. If $x_{i}\left(t\right)>a\left(T_{i}^{c}\right),\forall t\in\left(t_{1},t_{2}\right)$,
then $x_{i}\left(t\right)$ consumes more power to run extra servers
than $\bar{x_{i}}^{*}\left(t\right)$ in each period $\left(t_{1},t_{2}\right)$.
If $x_{i}\left(t\right)=a(t)=a\left(T_{i}^{c}\right)-1,\forall t\in\left(t_{1},t_{2}\right)$,
on one hand, $\bar{x_{i}}^{*}\left(t\right)$ costs at most $\left(T_{i+1}^{c}-T_{i}^{c}\right)P$
more power to run one extra server than $x_{i}\left(t\right)$ in
$\left(t_{1},t_{2}\right)$. On the other hand, $x_{i}\left(t\right)$
has to consume $\left(\beta_{on}+\beta_{off}\right)$ more power to
turn on/off a server one time in $\left(t_{1},t_{2}\right)$. Since
$\vartriangle\geq\left(T_{i+1}^{c}-T_{i}^{c}\right)$, we have $\left(\beta_{on}+\beta_{off}\right)\geq\left(T_{i+1}^{c}-T_{i}^{c}\right)P$.
This means $\bar{x_{i}}^{*}\left(t\right)$ does not cost more power
than $x_{i}\left(t\right)$ in $\left(t_{1},t_{2}\right)$. Therefore,
in both situations $\bar{x_{i}}^{*}\left(t\right)$ does not cost
more power than $x_{i}\left(t\right)$ in period $\left(t_{1},t_{2}\right)$.
If there exist other periods like $\left(t_{1},t_{2}\right)$,(One
example is $\left(t_{3},t_{4}\right)$ in Fig. \ref{fig:example2})
we can prove that $\bar{x_{i}}^{*}\left(t\right)$ does not cost more
power than $x_{i}\left(t\right)$ in these periods in the same way
as we did for $\left(t_{1},t_{2}\right)$. On the other hand, $x_{i}\left(t\right)$
and $\bar{x_{i}}^{*}\left(t\right)$ have the same power consumption
in the rest of time in $\left(T_{i}^{c},T_{i+1}^{c}\right)$. It follows
that $\bar{x_{i}}^{*}\left(t\right)$ does not cost more power than
$x_{i}\left(t\right)$ in $\left(T_{i}^{c},T_{i+1}^{c}\right)$, which
means $\bar{x_{i}}^{*}\left(t\right)$ is an optimal solution to \eqref{eq:sub}.

\begin{figure}
\begin{centering}
\includegraphics[width=8cm]{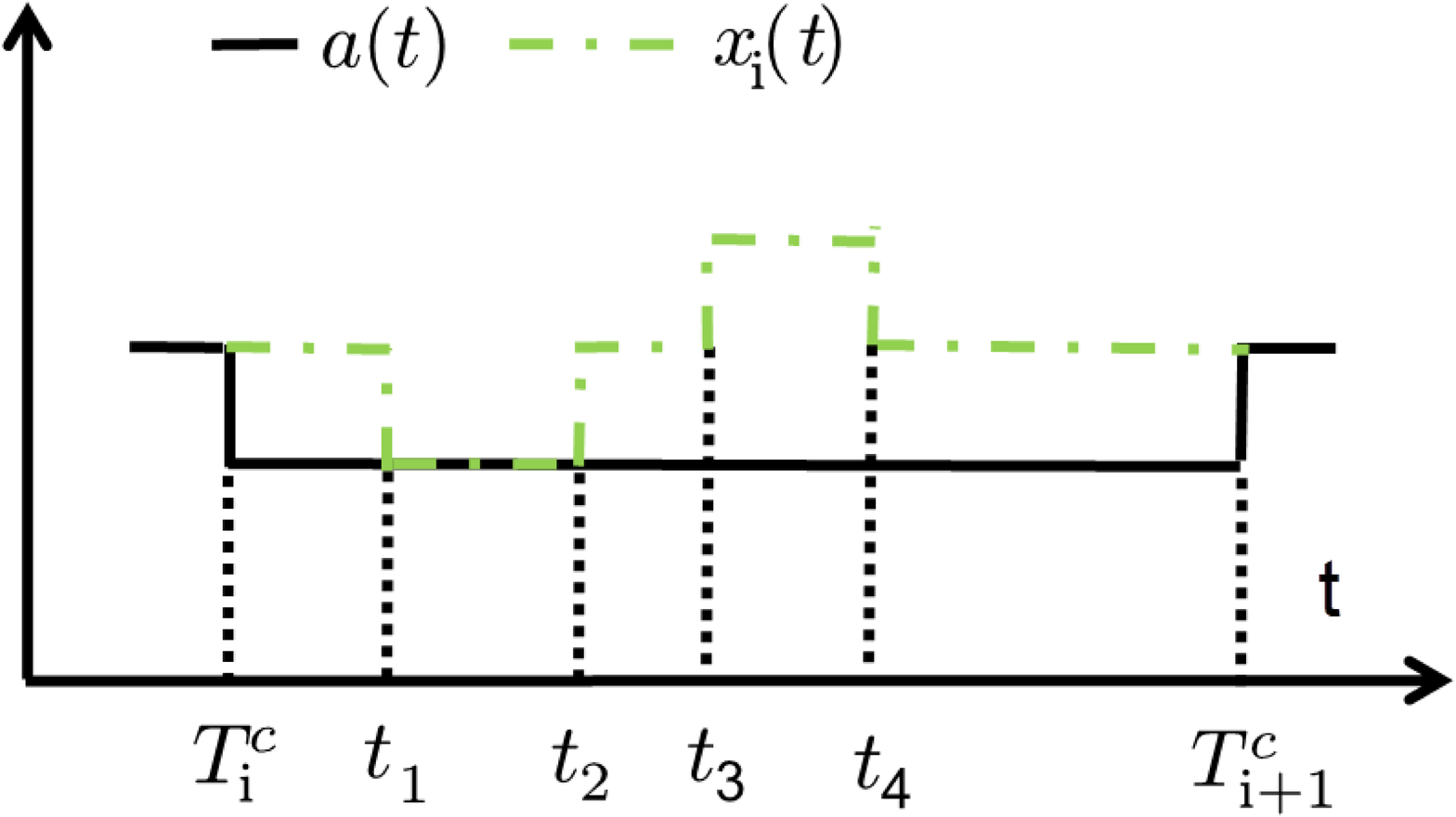}
\par\end{centering}

\caption{An example of solution $x_{i}\left(t\right)$ to \eqref{eq:sub} in
Type-III critical segment. $x_{i}\left(t\right)$ is not equal to
$a\left(T_{i}^{c}\right)$ in $\left(t_{1},t_{2}\right)$ and $\left(t_{3},t_{4}\right)$.\label{fig:example2}}

\end{figure}

Case 2: $\vartriangle<\left(T_{i+1}^{c}-T_{i}^{c}\right)$.

In this case, we claim that $\bar{x_{i}}^{*}\left(t\right)=a\left(T_{i}^{c}\right)-1,\forall t\in\left(T_{i}^{c},T_{i+1}^{c}\right)$
can achieve $P_{i}^{*}$. Because we can turn off the new idle server
at $T_{i}^{c}$ and turn on the server at $T_{i+1}^{c}$. In this
way, we can save $\left(T_{i+1}^{c}-T_{i}^{c}\right)P$ power consumption
which is greater then the on-off cost $\beta_{on}+\beta_{off}$. Thus,
$\bar{x_{i}}^{*}\left(t\right)=a\left(T_{i}^{c}\right)-1,\forall t\in\left(T_{i}^{c},T_{i+1}^{c}\right)$
can achieve $P_{i}^{*}$.

For Type-IV segment, we divide the situation in two cases in the same
way as we did for Type-III segment.

Case 1: $\vartriangle\geq\left(T_{i+1}^{c}-T_{i}^{c}\right)$.

In this case, we claim that $\bar{x_{i}}^{*}\left(t\right)=a\left(T_{i}^{c}\right),\forall t\in\left(T_{i}^{c},T_{i+1}^{c}\right)$
can achieve $P_{i}^{*}$. The proof is similar to the proof for Type-III
critical segment under the same situation $\vartriangle\geq\left(T_{i+1}^{c}-T_{i}^{c}\right)$.
Let $x_{i}\left(t\right)$ be any solution to \eqref{eq:sub} and
$x_{i}\left(t\right)$ is not always equal to $a\left(T_{i}^{c}\right)$
during $\left(T_{i}^{c},T_{i+1}^{c}\right)$. We will prove that $\bar{x_{i}}^{*}\left(t\right)$
does not cost more power than $x_{i}\left(t\right)$ in $\left(T_{i}^{c},T_{i+1}^{c}\right)$.
Because $x_{i}\left(t\right)$ is not always equal to $a\left(T_{i}^{c}\right)$
during $\left(T_{i}^{c},T_{i+1}^{c}\right)$, we can find period $\left(t_{1},t_{2}\right)\subseteq\left(T_{i}^{c},T_{i+1}^{c}\right)$
such that $a\left(T_{i}^{c}\right)=x_{i}\left(t_{1}\right)$, $a\left(T_{i}^{c}\right)=x_{i}\left(t_{2}\right)$
and $x_{i}\left(t\right)\neq a\left(T_{i}^{c}\right),\forall t\in\left(t_{1},t_{2}\right)$.
One example of such period is $\left(t_{1},t_{2}\right)$ in Fig.
\ref{fig:example3}.

First, we will compare the power consumed by $x_{i}\left(t\right)$
and $\bar{x_{i}}^{*}\left(t\right)$ in $\left(t_{1},t_{2}\right)$
based on two situations. If $x_{i}\left(t\right)>a\left(T_{i}^{c}\right),\forall t\in\left(t_{1},t_{2}\right)$,
then $x_{i}\left(t\right)$ consumes more power to run extra servers
than $\bar{x_{i}}^{*}\left(t\right)$ in period $\left(t_{1},t_{2}\right)$.
If $x_{i}\left(t\right)<a\left(T_{i}^{c}\right)-1,\forall t\in\left(t_{1},t_{2}\right)$,
which means a certain number of servers has been turned off during
$\left(t_{1},t_{2}\right)$ for certain amount of time. Denote $\gamma$
as the total number of servers have been turned off during $\left(t_{1},t_{2}\right)$.
On one hand, $\bar{x_{i}}^{*}\left(t\right)$ cost at most $\gamma\left(T_{i+1}^{c}-T_{i}^{c}\right)P$
power to run extra servers in $\left(t_{j},t_{j}^{'}\right)$. On
the other hand, $x_{i}\left(t\right)$ has to consume $\gamma_{j}\left(\beta_{on}+\beta_{off}\right)$
power to turn on/off servers $\gamma_{j}$ times in $\left(t_{1},t_{2}\right)$.
Since $\vartriangle\geq\left(T_{i+1}^{c}-T_{i}^{c}\right)$, we have
$\gamma\left(\beta_{on}+\beta_{off}\right)\geq\gamma\left(T_{i+1}^{c}-T_{i}^{c}\right)P$
. This means $\bar{x_{i}}^{*}\left(t\right)$ does not cost more power
than $x_{i}\left(t\right)$ in $\left(t_{1},t_{2}\right)$. Therefore,
in both situation $\bar{x_{i}}^{*}\left(t\right)$ does not cost more
power than $x_{i}\left(t\right)$ in each period $\left(t_{1},t_{2}\right)$.
Moreover, $x_{i}\left(t\right)$ and $\bar{x_{i}}^{*}\left(t\right)$
have the same power consumption in the rest of time in $\left(T_{i}^{c},T_{i+1}^{c}\right)$.
It follows that $\bar{x_{i}}^{*}\left(t\right)$ does not cost more
power than $x_{i}\left(t\right)$ in $\left(T_{i}^{c},T_{i+1}^{c}\right)$,
which means $\bar{x_{i}}^{*}\left(t\right)$ is an optimal solution
to \eqref{eq:sub}.

We consider Type-I, Type-II, Type-III and Type-IV segment with $\vartriangle\geq\left(T_{i+1}^{c}-T_{i}^{c}\right)$
to be the four basic critical segments, based on which we discuss
the case of Type-IV segment with $\vartriangle<\left(T_{i+1}^{c}-T_{i}^{c}\right)$.

\begin{figure}
\begin{centering}
\includegraphics[width=8cm]{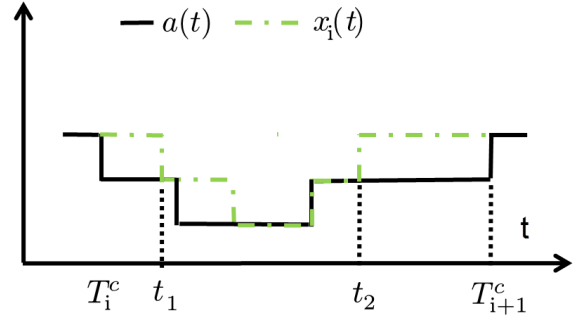}
\par\end{centering}

\caption{An example of solution $x_{i}\left(t\right)$ to \eqref{eq:sub} in
Type-IV critical segment. $x_{i}\left(t\right)$ is not equal to $a\left(T_{i}^{c}\right)$
in $\left(t_{1},t_{2}\right)$.\label{fig:example3} }

\end{figure}

Case 2: $\vartriangle<\left(T_{i+1}^{c}-T_{i}^{c}\right)$.

Each job-departure epoch $\tau$ in $\left[T_{i}^{c},T_{i+1}^{c}\right]$
has a corresponding job-arrival epoch $\tau^{'}$ in $\left[T_{i}^{c},T_{i+1}^{c}\right]$
such that $a\left(\tau\right)=a\left(\tau^{'}\right)$ and $a\left(t\right)<a\left(\tau\right),\forall t\in\left(\tau,\tau^{'}\right)$.
And we can find a set of job-departure and arrival epoch pairs $\left(\tau_{1},\tau_{1}^{'}\right)$,$\left(\tau_{2},\tau_{2}^{'}\right)$...$\left(\tau_{L},\tau_{L}^{'}\right)$
according to the procedure in \textbf{Optimal Solution} \textbf{Construction
Procedure }for Type-IV critical segment with $\vartriangle<\left(T_{i+1}^{c}-T_{i}^{c}\right)$.

In order to prove that \textbf{Optimal Solution} \textbf{Construction
Procedure} constructs an optimal solution to \eqref{eq:sub} in $\left[T_{i}^{c},T_{i+1}^{c}\right]$
with $\vartriangle<\left(T_{i+1}^{c}-T_{i}^{c}\right)$, we are going
to prove that an optimal solution $x^{*}\left(t\right)$ to \eqref{eq:sub}
must meet $a\left(t\right)$ at every job-departure $\tau$ and its
corresponding job-arrival epoch $\tau^{'}$ if $\tau\notin\left(\tau_{l},\tau_{l}^{'}\right),l=1,2,...L$.
Based on this fact, we can prove that \textbf{Optimal Solution} \textbf{Construction
Procedure} constructs an optimal solution.

It is clear that if $\tau\notin\left(\tau_{l},\tau_{l}^{'}\right),l=1,2,...L$,
then we must have $\tau^{'}\notin\left(\tau_{l},\tau_{l}^{'}\right),l=1,2,...L$.
Otherwise, if $\tau^{'}\in\left(\tau_{l},\tau_{l}^{'}\right)$ for
some $l\in\left\{ 1,2,..L\right\} $, we must have $a\left(\tau_{l}\right)<a\left(\tau^{'}\right)$
because we have $a\left(t\right)<a\left(\tau\right),\forall t\in\left(\tau,\tau^{'}\right)$
for job-departure and arrival epoch $\left(\tau,\tau^{'}\right)$.
On the other hand, we also must have $a\left(\tau^{'}\right)<a\left(\tau_{l}\right)$
because $\tau^{'}\in\left(\tau_{l},\tau_{l}^{'}\right)$. This is
a contradiction with previous conclusion $a\left(\tau_{l}\right)<a\left(\tau^{'}\right)$.
Hence, $\tau^{'}\notin\left(\tau_{l},\tau_{l}^{'}\right),l=1,2,...L$.

Now, we are going to prove that the necessary condition for $x^{*}\left(t\right)$
to achieve optimal power consumption in $\left[T_{i}^{c},T_{i+1}^{c}\right]$
is that $x^{*}\left(t\right)$ must meet $a\left(t\right)$ at every
job-arrival $\tau$ and its corresponding job-arrival epoch $\tau^{'}$
if $\tau\notin\left(\tau_{l},\tau_{l}^{'}\right),l=1,2,...L$.

It is clear the necessary condition is satisfied when $\left(\tau,\tau^{'}\right)=\left(T_{i}^{c},T_{i+1}^{c}\right)$.
On the other hand, for any job-arrival and departure epoch pair $\left(\tau,\tau^{'}\right)\neq\left(T_{i}^{c},T_{i+1}^{c}\right)$
, we can always find another job-arrival and departure epoch pair
$\left(\mu,\mu^{'}\right)$ covering $\left(\tau,\tau^{'}\right)$,
i.e., $\left(\tau,\tau^{'}\right)\subset\left(\mu,\mu^{'}\right)$
and $a\left(\mu\right)=a\left(\tau\right)+1$. Moreover, we must also
have $\left(\mu^{'}-\mu\right)>\vartriangle$. Because if $\left(\mu^{'}-\mu\right)\leq\vartriangle$,
then we must have $\left(\mu,\mu^{'}\right)\subseteq\left(\tau_{l},\tau_{l}^{'}\right)$
for some $l\in\left\{ 1,2,..L\right\} $. This means $\tau\in\left(\tau_{l},\tau_{l}^{'}\right)$
for some $l\in\left\{ 1,2,..L\right\} $, which is a contradiction
with $\tau\notin\left(\tau_{l},\tau_{l}^{'}\right),l=1,2,...L$.

Since $\bar{x_{i}}^{*}\left(t\right)$ achieves the optimal power
consumption in $\left[T_{i}^{c},T_{i+1}^{c}\right]$, then $x^{*}\left(t\right),t\in\left(\mu,\mu^{'}\right)$
must be an optimal solution to $\mathcal{P}\left(x^{*}\left(\mu\right),x^{*}\left(\mu^{'}\right),\mu,\mu^{'}\right)$
with $\mu^{'}-\mu>\vartriangle$. It follows that $x^{*}\left(t\right),t\in\left(\mu,\mu^{'}\right)$
must satisfy the necessary condition of $\mathcal{P}\left(A,B,T_{s},T_{e}\right)$
problem stated in Lemma \ref{lem:necessary}, hence, $x^{*}\left(t\right)\leq a\left(\mu\right)-1,t\in\left(\mu,\mu^{'}\right)$.
Because $a\left(\tau\right)=a\left(\tau^{'}\right)=a\left(\mu\right)-1$,
we must have $x^{*}\left(\tau\right)=a\left(\tau\right)$ and $x^{*}\left(\tau^{'}\right)=a\left(\tau^{'}\right)$.

Note that according to the necessary condition, if $L=0$, then $x^{*}\left(t\right)$
must meet $a\left(t\right)$ at every job-departure $\tau$ and its
corresponding job-arrival epoch $\tau^{'}$ in $\left[T_{i}^{c},T_{i+1}^{c}\right]$.

We are ready to prove that \textbf{Optimal Solution} \textbf{Construction
Procedure} constructs an optimal solution to \eqref{eq:sub} in $\left[T_{i}^{c},T_{i+1}^{c}\right]$
with $\vartriangle<\left(T_{i+1}^{c}-T_{i}^{c}\right)$. We prove
it based on two cases.

$\left(a\right)$ For all the job-arrival and departure epoch pairs
$\left(\tau,\tau^{'}\right)$, we have $\left(\tau^{'}-\tau\right)>\vartriangle$.

In this case, $x^{*}\left(t\right)$ must meet $a\left(t\right)$
at every job-departure epoch $\tau$ and job-arrival epoch $\tau^{'}$
in $\left[T_{i}^{c},T_{i+1}^{c}\right]$ according to necessary condition
we just proved. It is easy to verify that $a\left(t\right)$ between
two consecutive epoches (the two epoches can be one of following four
cases: both are arrival epoches, both are departure epoch, the first
one is arrival epoch and the other one is departure epoch, the first
one is departure epoch and the other one is arrival epoch) is one
of the following smaller basic critical segments: Type-I, Type-II,
Type-III with $\left(\tau^{'}-\tau\right)>\vartriangle$. As we already
proved that $x^{*}\left(t\right)=a\left(t\right)$ is an optimal solution
in these smaller basic critical segments. Therefore, we must have
$x^{*}\left(t\right)=a\left(t\right),\forall t\left[T_{i}^{c},T_{i+1}^{c}\right]$,
which is the same as the solution $\bar{x_{i}}^{*}\left(t\right)$
constructed by our \textbf{Optimal Solution} \textbf{Construction
Procedure}. Hence, $\bar{x_{i}}^{*}\left(t\right)$ can achieve optimal
power consumption in $\left[T_{i}^{c},T_{i+1}^{c}\right]$.

$\left(b\right)$ There exist job-arrival and departure epoch pairs
$\left(\tau,\tau^{'}\right)$ such that $\left(\tau^{'}-\tau\right)\leq\vartriangle$.

In this case, $x^{*}\left(t\right)$ must meet $a\left(t\right)$
at all job-departure epoch $\tau$ and job-arrival epoch $\tau^{'}$
which are not in $\left(\tau_{1},\tau_{1}^{'}\right)\cup\left(\tau_{2},\tau_{2}^{'}\right)\cup...\cup\left(\tau_{L},\tau_{L}^{'}\right)$.
We also can verify that $a\left(t\right)$ in two consecutive epoches
which are not in $\left(\tau_{1},\tau_{1}^{'}\right)\cup\left(\tau_{2},\tau_{2}^{'}\right)\cup...\cup\left(\tau_{L},\tau_{L}^{'}\right)$
is one of the following smaller basic critical segments: Type-I, Type-II,
Type-III and Type-IV with $\left(\tau^{'}-\tau\right)\leq\vartriangle$.
Therefore, according to the optimal solution construction procedure
of the four basic critical segments, we must have $x^{*}\left(t\right)=a\left(\tau\right),\forall t\in\left(\tau,\tau^{'}\right)$
when $\left(\tau,\tau^{'}\right)$ is smaller basic critical segments
of Type-III with $\left(\tau^{'}-\tau\right)\leq\vartriangle$ and
Type-IV with $\left(\tau^{'}-\tau\right)\leq\vartriangle$. And in
the rest smaller basic segments, we must have $x^{*}\left(t\right)=a\left(t\right)$.
The whole $x^{*}\left(t\right),\forall t\in\left[T_{i}^{c},T_{i+1}^{c}\right]$
is the same as the solution $\bar{x_{i}}^{*}\left(t\right)$ constructed
by our \textbf{Optimal Solution} \textbf{Construction Procedure}.
Hence, $\bar{x_{i}}^{*}\left(t\right)$ can achieve optimal power
consumption in $\left[T_{i}^{c},T_{i+1}^{c}\right]$.

The above two cases cover all the possibility of $\left[T_{i}^{c},T_{i+1}^{c}\right]$.
We proved that in each case the solution constructed by \textbf{Optimal
Solution} \textbf{Construction Procedure} can achieve the optimal.
It follows that the solution constructed by \textbf{Optimal Solution}
\textbf{Construction Procedure} can achieve the optimal in $\left[T_{i}^{c},T_{i+1}^{c}\right]$
with $\vartriangle<\left(T_{i+1}^{c}-T_{i}^{c}\right)$.

Because we only have finite job arrival/departure in $\left[0,T\right]$
and each basic critical segment or smaller basic critical segment
contains at least one job arrival or departure epoch. Therefore, the
number of basic critical segments or smaller basic critical segments
is finite, which means our construction can terminate in finite time.

It is easy to verify that $x\left(t\right)$ constructed for critical
segments can connect to each other seamlessly. On the other hand,
the constructed $x\left(t\right)$ can achieve $P_{i}^{*}$ in each
critical segment, then the whole $x\left(t\right)$ can achieve the
lower bound of \eqref{eq: obj}, which means it is an optimal solution
to \eqref{eq: obj}. We have thus proved theorem \ref{Thm:opt_sol_pro_is_opt}.\end{IEEEproof}

\subsection{Proof of Theorem \ref{thm: offline}\label{apx:proof_5}}
\begin{IEEEproof}
First, we want to prove that the number of running servers $x_{o}\left(t\right)$
proposed by our off-line algorithm meets $a\left(t\right)$ at every
critical time $T_{i}^{c}$. We have $x_{o}\left(T_{1}^{c}\right)=a\left(T_{1}^{c}\right)$.
Given $x_{o}\left(T_{i}^{c}\right)=a\left(T_{i}^{c}\right)$, we want
to show that $x_{o}\left(T_{i+1}^{c}\right)=a\left(T_{i+1}^{c}\right)$.

$\left(a\right)$ if $T_{i}^{c}$ is an arrival epoch, then $\left[T_{i}^{c},T_{i+1}^{c}\right]$
is Type-I segment and there is no job departure during this critical
segment $\left(T_{i}^{c},T_{i+1}^{c}\right)$ and no idle server at
$T_{i}^{c}$. \textbf{Job-dispatching entity} just pops server ID
and turn on corresponding server to serve new job. Thus, we have $x_{o}\left(T_{i+1}^{c}\right)=a\left(T_{i+1}^{c}\right)$.

$\left(b\right)$ if $T_{i}^{c}$ is a departure epoch, then $\left[T_{i}^{c},T_{i+1}^{c}\right]$
is one of the rest three types critical segments and we must have
$a\left(T_{i+1}^{c}\right)=a\left(T_{i}^{c}\right)$ or $a\left(T_{i+1}^{c}\right)=a\left(T_{i}^{c}\right)-1$
.

When $a\left(T_{i+1}^{c}\right)=a\left(T_{i}^{c}\right)-1$ , the
system only has one idle server right after $T_{i}^{c}$ . The idle
server should make decision to remain idle or turn off. According
to the definition of $T_{i+1}^{c}$, the idle server can not find
arrival epoch $\tau$ after $T_{i}^{c}$ so that $a\left(\tau\right)=a\left(T_{i}^{c}\right)$.
Based on our off-line algorithm, the server will turn itself off.
Therefore, we have $x_{o}\left(T_{i+1}^{c}\right)=a\left(T_{i+1}^{c}\right)$.

When $a\left(T_{i+1}^{c}\right)=a\left(T_{i}^{c}\right)$, then the
critical segment is Type-III or Type-IV segment and $a\left(t\right)\leq a\left(T_{i}^{c}\right)-1,\forall t\in\left(T_{i}^{c},T_{i+1}^{c}\right)$.
Because the number of arrival epoches is less than the number of departure
epoches in $\left[T_{i}^{c},\tau\right],\forall\tau\in\left(T_{i}^{c},T_{i+1}^{c}\right)$
, which means \textbf{job-dispatching entity} pushed more server IDs
than popped in the period $\left[T_{i}^{c},\tau\right]$. Therefore,
\textbf{job-dispatching entity} will not pop server IDs pushed before
$T_{i}^{c}$ during $\left[T_{i}^{c},T_{i+1}^{c}\right]$, which means
the number of running servers during $\left[T_{i}^{c},T_{i+1}^{c}\right]$
is less than or equal to $x_{o}\left(T_{i}^{c}\right)$. Because we
have $x_{o}\left(T_{i}^{c}\right)=a\left(T_{i}^{c}\right)=a\left(T_{i+1}^{c}\right)$
and $x_{o}\left(T_{i}^{c}\right)\geq a\left(T_{i+1}^{c}\right)$ ,
we must have $a\left(T_{i+1}^{c}\right)=x_{o}\left(T_{i+1}^{c}\right)$.

By induction, we proved that $x_{o}\left(t\right)$ meets $a\left(t\right)$
at all the critical times.

Next, we are going to prove that $x_{o}\left(t\right)$ and the optimal
solution $x^{*}\left(t\right)$ constructed by \textbf{Optimal Solution}
\textbf{Construction Procedure} are the same. We divide the situation
into four cases.

Case 1: For Type-I segment $\left[T_{i}^{c},T_{i+1}^{c}\right]$.

Because there is no job departure during the non-decreasing critical
segment $\left(T_{i}^{c},T_{i+1}^{c}\right]$ and we have $a\left(T_{i}^{c}\right)=x_{o}\left(T_{i}^{c}\right)$,
which means there is on idle server at $T_{i}^{c}$ . According to
our off-line algorithm, \textbf{job-dispatching entity} just pops
server ID and turns on the corresponding server when new job arriving.
Thus, we have $x_{o}\left(t\right)=a\left(t\right)=x^{*}\left(t\right),\forall t\in\left[T_{i}^{c},T_{i+1}^{c}\right]$.

Case 2: For Type-II segment $\left[T_{i}^{c},T_{i+1}^{c}\right]$.

According to proposition \ref{prop:property}, for step-decreasing
segment we have $a\left(t\right)=a\left(T_{i}^{c}\right)-1,\forall t\in\left(T_{i}^{c},T_{i+1}^{c}\right)$.
After job departure at $T_{i}^{c}$, the new idle corresponding server
can not find time $t_{1}\in(T_{i}^{c},T_{i}^{c}+\Delta]$ so that
$a(t_{1})=a(T_{i}^{c})$. Hence, based on our off-line algorithm,
the server turns itself off and we have $x_{o}\left(t\right)=a\left(T_{i}^{c}\right)-1=a\left(t\right)=x^{*}\left(t\right),\forall t\in\left[T_{i}^{c},T_{i+1}^{c}\right]$.

Case 3: For Type-III segment $\left[T_{i}^{c},T_{i+1}^{c}\right]$.

For Type-III segment, \textbf{job-dispatching entity }will push a
server ID at $T_{i}^{c}$ and pop it at $T_{i+1}^{c}.$ If $\vartriangle<\left(T_{i+1}^{c}-T_{i}^{c}\right)$,
the corresponding server can not find time $t_{1}\in(T_{i}^{c},T_{i}^{c}+\Delta]$
so that $a(t_{1})=a(T_{i}^{c})$, our off-line algorithm will turn
off the corresponding server and $x_{o}\left(t\right)=a\left(T_{i}^{c}\right)-1,\forall t\in\left[T_{i}^{c},T_{i+1}^{c}\right]$.
If $\vartriangle\geq\left(T_{i+1}^{c}-T_{i}^{c}\right)$, the server
will remain idle and $x_{o}\left(t\right)=a\left(T_{i}^{c}\right),\forall t\in\left[T_{i}^{c},T_{i+1}^{c}\right]$.
Hence, we have $x_{o}\left(t\right)=x^{*}\left(t\right),\forall t\in\left[T_{i}^{c},T_{i+1}^{c}\right]$.

Case 4: For Type-IV segment $\left[T_{i}^{c},T_{i+1}^{c}\right]$.

In this case, if $\vartriangle\geq\left(T_{i+1}^{c}-T_{i}^{c}\right)$,
at each departure epoch in $\left[T_{i}^{c},T_{i+1}^{c}\right]$,
the corresponding new idle server can find time $t_{2}\in(t_{1},t_{1}+\Delta]$
so that $a(t_{2})=a(t_{1})$, where $t_{1}$ is the departure epoch.
Therefore, all the servers remain idle according to our off-line algorithm
and $x_{o}\left(t\right)=a\left(T_{i}^{c}\right),\forall t\in\left[T_{i}^{c},T_{i+1}^{c}\right]$.
If $\vartriangle\leq\left(T_{i+1}^{c}-T_{i}^{c}\right)$, at each
departure epoch $\tau$, our offline algorithm will turn off the new
idle server if the corresponding departure epoch $\tau^{'}$ satisfying
that $\tau^{'}-\tau>\vartriangle$ because the idle server can not
find time $t_{1}\in(\tau,\tau+\Delta]$ so that $a(\tau)=a(t_{1})$.
If $\tau^{'}-\tau\leq\vartriangle$, the new idle server will remain
idle. In this way, the number of running servers $x_{o}\left(t\right)$
decided by our off-line algorithm is equal to $x_{o}\left(t\right)=x^{*}\left(t\right),\forall t\in\left[T_{i}^{c},T_{i+1}^{c}\right]$.

Based on above four cases, we prove that $x_{o}\left(t\right)=x^{*}\left(t\right)$.
Therefore, it can achieve the optimal value for \eqref{eq: obj} in
offline situation according to theorem \ref{Thm:opt_sol_pro_is_opt}.\end{IEEEproof}

\subsection{Proof of Theorem \ref{thm:online}\label{apx:proof_6}}

We are going to prove theorem \ref{thm:online}. Before doing so,
we first prove Lemma \ref{lem:For-the-same} and two other lemmas.

\noindent\textbf{Lemma \ref{lem:For-the-same}}: \emph{For the same
$a\left(t\right),t\in\left[0,T\right]$, under the last-empty-server-first
job-dispatching strategy, each server will get the same job at the
same time and the job will leave the server at the same time for both
off-line and online situations.}
\begin{IEEEproof}
For both off-line and online situation, we have the same $a\left(0\right)$
servers running at $t=0$. The other servers are off and their IDs
are stored in the stack in the same order at $t=0$. Let $\Gamma_{i}$
denote the $i$th epoch that a job departs or arrivals the system
in $\left[0,T\right]$. Assume the number of total arrival and departure
epoches is $S$. To prove Lemma \ref{lem:For-the-same}, we first
claim that same server IDs are stored in the stack in the same order
for both off-line and online situation in each period $\left[\Gamma_{i},\Gamma_{i+1}\right),i=1,2,...S-1$.
Moreover, both situations have the same servers running and each running
server serve the same corresponding job in $\left[\Gamma_{i},\Gamma_{i+1}\right)$.
We will prove the claim by induction.

First, we prove that the claim is true for $\left[\Gamma_{1},\Gamma_{2}\right)$.
If $\Gamma_{1}$ is a job-arrival epoch, for both off-line situation
and online situation, the \textbf{job-dispatching entity }will pop
the same server ID to server the new job because both off-line and
online situation have the same server IDs in stack and IDs are in
the same order at $t=0$. After popping the server ID at the top of
the stack at $\Gamma_{1}$, both off-line and online situation still
have the same server IDs stored in the stack and IDs are in the same
order. And both situations have the same servers running and each
running server serve the same job. Because there is no job arrival
or departure in $\left(\Gamma_{1},\Gamma_{2}\right)$, therefore,
no server ID will be popped out of the stack or pushed in the stack
during $\left(\Gamma_{1},\Gamma_{2}\right)$, which means both the
two situations will remain having the same server IDs stored in the
stack in the same order and having the same servers running. Moreover,
each running server serve the same corresponding job during $\left(\Gamma_{1},\Gamma_{2}\right)$
in both two situations.

If $\Gamma_{1}$ is a job-departure epoch, for both off-line situation
and online situation, the \textbf{job-dispatching entity }will push
the same server ID in the stack because both off-line and online situation
have the same servers serving the same jobs at $t=0$. After pushing
the server ID in the stack at $\Gamma_{1}$, both off-line and online
situation still have the same server IDs stored in the stack and IDs
are in the same order, Moreover, both situations have the same servers
running and each running server serve the same job. Because there
is no job arrival or departure in $\left(\Gamma_{1},\Gamma_{2}\right)$,
both the two situations will remain having the same server IDs stored
in the stack in the same order and having the same servers running
and each running server serve the same job during $\left(\Gamma_{1},\Gamma_{2}\right)$.
Therefore, the claim is true for $\left[\Gamma_{1},\Gamma_{2}\right)$
no matter $\Gamma_{1}$is a job-arrival or departure epoch.

Next, we will prove that same server IDs are stored in the stack in
the same order for both off-line and online situation in period $\left[\Gamma_{i},\Gamma_{i+1}\right)$,
Moreover, both off-line and online situation have the same serving
running and each running server serve the same job in two situations
in period $\left[\Gamma_{i},\Gamma_{i+1}\right)$, given that both
the two situations have the same server IDs stored in the stack in
the same order and both off-line and online situation have the same
serving running and each running server serve the same job in two
situations in $\left[\Gamma_{i-1},\Gamma_{i}\right)$. The proof is
also based on two cases. If $\Gamma_{i}$ is a job-arrival epoch,
for both off-line situation and online situation, the \textbf{job-dispatching
entity }will pop the same server ID to server the new job because
both off-line and online situation have the same server IDs in stack
and IDs are in the same order in $\left[\Gamma_{i-1},\Gamma_{i}\right)$.
After popping the server ID at the top of the stack at $\Gamma_{i}$,
both off-line and online situation still have the same server IDs
stored in the stack and IDs are in the same order. They also have
the same running servers and each server server the same job due to
both the situation have the same servers running and each server serve
the same job in $\left[\Gamma_{i-1},\Gamma_{i}\right)$. Because there
is no job arrival or departure in $\left(\Gamma_{i},\Gamma_{i+1}\right)$,
therefore, no server ID will be popped out of the stack or pushed
in the stack during $\left(\Gamma_{i},\Gamma_{i+1}\right)$, which
means both the two situations will remain having the same server IDs
stored in the stack in the same order and having the same servers
running and each running server serve the same job during $\left(\Gamma_{i},\Gamma_{i+1}\right)$.

If $\Gamma_{i}$ is a job-departure epoch, for both off-line situation
and online situation, the \textbf{job-dispatching entity }will push
the same server ID in the stack because both off-line and online situation
have the same servers serving the same jobs in $\left[\Gamma_{i-1},\Gamma_{i}\right)$.
After pushing the server ID in the stack at $\Gamma_{i}$, both off-line
and online situation still have the same server IDs stored in the
stack and IDs are in the same order. Moreover, They also have the
same running servers and each server server the same job due to both
the situation have the same servers running and each server serve
the same job in $\left[\Gamma_{i-1},\Gamma_{i}\right)$. Because there
is no job arrival or departure in $\left(\Gamma_{i},\Gamma_{i+1}\right)$,
both the two situations will remain having the same server IDs stored
in the stack in the same order and having the same servers running
and each running server serve the same job during $\left(\Gamma_{i},\Gamma_{i+1}\right)$.
Therefore, the claim is true for $\left[\Gamma_{i},\Gamma_{i+1}\right)$
no matter $\Gamma_{i}$ is a job-arrival or departure epoch.

Up to now, we proved that same server IDs are stored in the stack
in the same order for both off-line and online situation in each period
$\left[\Gamma_{i},\Gamma_{i+1}\right),i=1,2,...S-1$. Moreover, both
situations have the same servers running and each running server serve
the same job in $\left[\Gamma_{i},\Gamma_{i+1}\right)$. Due to this
fact, we can prove Lemma \ref{lem:For-the-same}. If a server get
a job at a job-arrival epoch in online situation, then same server
will get the same job at the job-arrival epoch in off-line situation
because both the situation have same server IDs stored on the top
of the stack. On the other hand, if a job leave a server in online
situation, then the same job will leave the same server because both
situation have the same running server to serve the same job.\end{IEEEproof}
\begin{lem}
The deterministic online ski-rental algorithm we applied in our online
algorithm \textbf{A1} has competitive ratio $2-\alpha$.\label{lem:The-deterministic-online}\end{lem}
\begin{IEEEproof}
As we already proved in Lemma \ref{lem:For-the-same}, for both online
and off-line cases, a server faces the same set of jobs. From now
on, we focus on one server. \textbf{Job-dispatching entity }will assign
job to the server form time to time and we assume that the server
will serve total $W$ jobs in $\left[0,T\right]$. Denote $\tau_{j,s}$
as the time in $\left[0,T\right]$ that the server gets its $j$th
job and define $\tau_{j,e}$ as the time that $j$th job of the server
leaves the system. Define $\tau_{W+1,s}=T$. The server should decide
to turn off itself of stay idle between $\tau_{j,e}$ and $\tau_{j+1,s}$.
In order to get competitive ratio of the deterministic online ski-rental
algorithm we applied in \textbf{A1}, we want to compare the power
consumption $P_{j,on}$ of the online ski-rental algorithm in $\left(\tau_{j,s},\tau_{j+1,s}\right],j\leq W$
with the power consumption $P_{j,off}$ of off-line ski-rental algorithm
in $\left(\tau_{j,s},\tau_{j+1,s}\right]$. In fact, the power consumption
of the online and off-line ski-rental algorithms depend on the length
of the time between $\tau_{j,e}$ and $\tau_{j+1,s}$. Denote $T_{j,B}=\left(T_{j,e}-T_{j,s}\right)$
as the length of busy period in $\left(\tau_{j,s},\tau_{j+1,s}\right]$
and $T_{j,E}=\left(T_{j+1,s}-T_{j,e}\right)$ as the length of empty
period in $\left(\tau_{j,s},\tau_{j+1,s}\right]$, then we have:

\begin{equation}
P_{j,off}=\begin{cases}
PT_{j,B}+PT_{j,E} & if\, T_{j,E}\leq\vartriangle\\
PT_{j,B}+\left(\beta_{on}+\beta_{off}\right) & if\, T_{j,E}>\vartriangle
\end{cases}
\end{equation}

According to the online ski-rental algorithm in \textbf{A1}, we also
have:

\begin{equation}
P_{j,on}=\begin{cases}
PT_{j,B}+PT_{j,E} & if\, T_{j,E}\leq\vartriangle\\
PT_{j,B}+\left(\beta_{on}+\beta_{off}\right)+P\left(1-\alpha\right)\vartriangle & if\, T_{j,E}>\vartriangle
\end{cases}
\end{equation}

Hence, when $T_{j,E}\leq\vartriangle$, $\frac{P_{j,on}}{P_{j,off}}=1$,
when $T_{j,E}>\vartriangle$

\begin{eqnarray*}
\frac{P_{j,on}}{P_{j,off}} & =\frac{PT_{j,B}+\left(\beta_{on}+\beta_{off}\right)+P\left(1-\alpha\right)\vartriangle}{PT_{j,B}+\left(\beta_{on}+\beta_{off}\right)}\\
 & \leq\frac{\left(\beta_{on}+\beta_{off}\right)+P\left(1-\alpha\right)\vartriangle}{\left(\beta_{on}+\beta_{off}\right)}=2-\alpha
\end{eqnarray*}

In the above calculation, we used $P\vartriangle=\left(\beta_{on}+\beta_{off}\right)$
and we have $\frac{P_{j,on}}{P_{j,off}}\leq2-\alpha,\alpha\in\left[0,1\right]$
for any $T_{j,E}$. On the other hand, for any $j=1,2,\ldots W$,
we have $\frac{P_{j,on}}{P_{j,off}}\leq2-\alpha,\alpha\in\left[0,1\right]$.
Therefore, the power consumption of the online ski-rental algorithm
in $\left[0,T\right]$ is at most $\left(2-\alpha\right)$ times the
optimal, which means the competitive ratio of the deterministic online
ski-rental algorithm applied in \textbf{A1} is $2-\alpha$. \end{IEEEproof}
\begin{lem}
The randomized online ski-rental algorithm we applied in our online
algorithm \textbf{A2} has competitive ratio $\left(e-\alpha\right)/\left(e-1\right)$.\label{lem:The-randomized-online}\end{lem}
\begin{IEEEproof}
In the proof, we still focus on one server. we will use the same notations
we used to prove Lemma \ref{lem:The-deterministic-online}. We want
to compare the average power consumption $P_{j,on}$ of the randomized
online ski-rental algorithm in $\left(\tau_{j,s},\tau_{j+1,s}\right],j\leq W$
with power consumption $P_{j,off}$ of off-line ski-rental algorithm
in $\left(\tau_{j,s},\tau_{j+1,s}\right]$. we have:

\begin{equation}
P_{j,off}=\begin{cases}
PT_{j,B}+PT_{j,E}, & T_{j,E}\leq\vartriangle\\
PT_{j,B}+\left(\beta_{on}+\beta_{off}\right), & T_{j,E}>\vartriangle
\end{cases}
\end{equation}

And according to the randomized online ski-rental algorithm, when
$T<\alpha\vartriangle$, we have
\[
E\left(P_{j,on}\right)=PT_{j,B}+PT_{j,E}.
\]
When $\alpha\vartriangle\leq T_{j,E}\leq\vartriangle$, we have
\begin{eqnarray*}
E\left(P_{j,on}\right) & = & PT_{j,B}+P\varint_{0}^{T_{j,E}-\alpha\vartriangle}\left(z+\beta_{on}+\beta_{off}\right)f_{Z}\left(z\right)dz\\
 &  & +P\varint_{T_{j,E}-\alpha\vartriangle}^{\left(1-\alpha\right)\vartriangle}T_{j,E}f_{Z}\left(z\right)dz.
\end{eqnarray*}
When $T_{j,E}>\vartriangle,$ we have
\[
E\left(P_{j,on}\right)=PT_{j,B}+P\varint_{0}^{\left(1-\alpha\right)\vartriangle}\left(z+\beta_{on}+\beta_{off}\right)f_{Z}\left(z\right)dz.
\]

We get the above expected power consumption for $\alpha\vartriangle\leq T_{j,E}\leq\vartriangle$
based on following reason: If the number $Z$ generated by the server
is less than $T_{j,E}-\alpha\vartriangle$, then the server will waits
for $Z$ amount of time, consuming $PZ$ power. And it looks into
the prediction window of size $\alpha\Delta$ and find it won't receive
any job during the window because $Z<T_{j,E}-\alpha\vartriangle$.
Therefore, it turns itself off and cost power $\left(\beta_{on}+\beta_{off}\right)$.
On the other hand, if $Z\geq T_{j,E}-\alpha\vartriangle$, the server
will not turn itself off and consume $PT_{j,E}$ to stay idle. We
can get the expected power consumption for $T_{j,E}<\alpha\vartriangle$
and $T_{j,E}>\vartriangle$ in the same way. Because

\[
f_{Z}(z)=\begin{cases}
\frac{e^{z/\left(1-\alpha\right)\Delta}}{\left(e-1\right)\left(1-\alpha\right)\Delta}, & \mbox{if }0\leq z\leq\left(1-\alpha\right)\Delta;\\
0, & \mbox{otherwise.}
\end{cases}
\]

We can calculate $E\left(P_{j,on}\right)$ and the ratio between $E\left(P_{j,on}\right)$
and $P_{j,off}$:

$\frac{E\left(P_{j,on}\right)}{P_{j,off}}=\begin{cases}
1, & T_{j,E}<\alpha\vartriangle\\
\frac{e}{e-1}-\frac{\alpha}{e-1}\frac{\vartriangle}{T_{j,E}}, & \alpha\vartriangle\leq T_{j,E}\leq\vartriangle\\
\frac{e-\alpha}{e-1} & T_{j,E}>\vartriangle
\end{cases}$

From above expression, we can conclude that $\frac{E\left(P_{j,on}\right)}{P_{j,off}}\leq\frac{e-\alpha}{e-1}$
for any $T_{j,E}$. On the other hand, for any $j=1,2,\ldots W$,
we have $\frac{E\left(P_{j,on}\right)}{P_{j,off}}\leq\frac{e-\alpha}{e-1},\alpha\in\left[0,1\right]$.
Therefore, the power consumption of the online ski-rental algorithm
in $\left[0,T\right]$ is at most $\frac{e-\alpha}{e-1}$ times the
optimal, which means the competitive ratio of the randomized online
ski-rental algorithm applied in \textbf{A2} is $\frac{e-\alpha}{e-1}$. \end{IEEEproof}
\begin{lem}
The randomized online ski-rental algorithm we applied in our online
algorithm \textbf{A3} has competitive ratio $e/\left(e-1+\alpha\right)$.\label{lem:Best-randomized-online}\end{lem}
\begin{IEEEproof}
The only difference between \textbf{A2} and \textbf{A3} is that the
random variable $Z$ has different probability distribution. Therefore,
the proof of Lemma \ref{lem:Best-randomized-online} is the same as
the proof of Lemma \ref{lem:The-randomized-online}. And it can be
easily verified that \textbf{A3} has competitive ratio $e/\left(e-1+\alpha\right)$
\end{IEEEproof}
Now we are ready to prove theorem \textit{\ref{thm:online}.}
\begin{IEEEproof}
As we already proved in our off-line algorithm that the optimal power
consumption of the data center can be achieved by each server run
off-line ski-rental algorithm individually and independently. On the
other hand, in Lemma \ref{lem:The-deterministic-online}, \ref{lem:The-randomized-online}
and \ref{lem:Best-randomized-online}, we proved that the power consumption
of deterministic and randomized online ski-rental algorithm we applied
are at most $2-\alpha$, $\frac{e-\alpha}{e-1}$ and $\frac{e}{e-1+\alpha}$
times the power consumption of off-line ski-rental algorithm for one
server. Therefore, the power consumption of our online algorithm \textbf{A1},
\textbf{A2} and \textbf{A3} are at most $2-\alpha$, $\frac{e-\alpha}{e-1}$
and $\frac{e}{e-1+\alpha}$ times the power consumption of off-line
algorithm for data center, which means the competitive ratios of \textbf{A1},
\textbf{A2} and \textbf{A3} are $2-\alpha$, $\frac{e-\alpha}{e-1}$
and $\frac{e}{e-1+\alpha}$ respectively.

Next, we want to prove that \textbf{A1 }has the best competitive ratio
for deterministic online algorithms under our job-dispatching strategy.
In fact, assume that deterministic online algorithm peeks into the
future window and then decide to turn off itself or stay idle $\theta\vartriangle$
after becoming empty at $t_{1}$. When $\theta<1-\alpha$, if the
server will receive its next job right after $t_{1}+\left(\theta+\alpha\right)\vartriangle$,
then the online algorithm will turn off itself at $t_{1}+\theta\vartriangle$,
and consume $P\left(\theta+1\right)\vartriangle$ power. On the other
hand, the offline optimal is $\left(\alpha+\theta\right)P\vartriangle$.
The competitive ratio is at least $\frac{1+\theta}{\theta+\alpha}>2-\alpha$.

When $\theta>1-\alpha$, if the server will receive its next job right
after $t_{1}+\left(\theta+\alpha\right)\vartriangle$, then the online
algorithm will turn off itself at $t_{1}+\theta\vartriangle$, and
consume $P\left(\theta+1\right)\vartriangle$ power. On the other
hand, the offline optimal is $P\vartriangle$. The competitive ratio
at least is $1+\theta>2-\alpha$.

Based on above two cases, we can see that only when $\theta=1-\alpha$,
the deterministic algorithm has better competitive ratio $2-\alpha$.
Therefore, the best deterministic online algorithm is \textbf{A1},
which has competitive ratio $2-\alpha$.

Finally, we want to prove that \textbf{A3 }has the best competitive
ratio for randomized online algorithms under our job-dispatching strategy.
In fact, assume that the server becomes empty at $\tau_{1}$ and it
will receive its next job at $\tau_{2}$. In order to find the best
competitive ratio for randomized online algorithm, according to the
proof of Lemma \ref{lem:The-randomized-online}, it is sufficient
to find the minimal ratio of the power consumed by randomized online
algorithm to that of the offline optimal in $\left[\tau_{1},\tau_{2}\right]$.
We first chop time period $\left(\tau_{1},\tau_{2}\right)$ into small
time slot. Then we let the length of slot goes to zero, we can get
the best competitive ratio for continuous time randomized online algorithm.

Assume critical interval $\vartriangle$ contains exact $b$ slots
and there are $D$ slots in $\left[\tau_{1},\tau_{2}\right]$. Moreover,
the future window has $k\leq b-1$ slots. (If $k\geq b$, the online
algorithm can achieve the offline optimal and the competitive ratio
is 1.) Let $P_{i}$ denote the probability that the algorithm decides
to turn off the server at slot $i$. Define $c$ as the competitive
ratio. Then we can solve following optimization problem to find the
minimal competitive ratio.

\begin{eqnarray}
 & \inf & c\label{eq: BCR1}\\
 & \textrm{s.t}. & D\overset{\infty}{\underset{i=1}{\sum}}P_{i}\leq cD,\forall D\in[0,k],\label{eq:-2}\\
 &  & \overset{D-k}{\underset{i=1}{\sum}}\left(b+i-1\right)P_{i}+\overset{\infty}{\underset{i=D-k+1}{\sum}}DP_{i}\leq Dc,\forall D\in\left(k,b\right]\label{eq:}\\
 &  & \overset{D-k}{\underset{i=1}{\sum}}\left(b+i-1\right)P_{i}+\overset{\infty}{\underset{i=D-k+1}{\sum}}DP_{i}\leq bc,\forall D\in\left(b,\infty\right]\\
 &  & \overset{b-k}{\underset{i=1}{\sum}}P_{i}=1\\
 & \textrm{var} & c,P_{i},\forall i\in\left\{ 1,2,\ldots,\infty\right\}
\end{eqnarray}

We are going to prove that the optimal value $c^{*}$ of problem \eqref{eq: BCR1}
is equal to the optimal value $\bar{c}^{*}$ of following problem.

\begin{eqnarray}
 & \textrm{min} & \bar{c}\label{eq: BCR2}\\
 & \textrm{s.t}. & \bar{D}\overset{b-k}{\underset{i=1}{\sum}}\bar{P}_{i}\leq\bar{c}\bar{D},\forall\bar{D}\in[0,k],\label{eq:-2-1}\\
 &  & \overset{\bar{D}-k}{\underset{i=1}{\sum}}\left(b+i-1\right)\bar{P}_{i}+\overset{b-k}{\underset{i=\bar{D}-k+1}{\sum}}\bar{D}\bar{P}_{i}\leq\bar{D}\bar{c},\forall\bar{D}\in\left(k,b\right)\label{eq:-1}\\
 &  & \overset{b-k}{\underset{i=1}{\sum}}\left(b+i-1\right)\bar{P}_{i}\leq b\bar{c},\forall\bar{D}\in\left[b,\infty\right]\\
 &  & \overset{b-k}{\underset{i=1}{\sum}}\bar{P}_{i}=1\\
 & \textrm{var} & \bar{c},\bar{P_{i}},\forall i\in\left\{ 1,2,\ldots,b-k\right\}
\end{eqnarray}

First, it is easy to see that every solution to \eqref{eq: BCR2}
is a solution to \eqref{eq: BCR1}. Therefore, we have $c^{*}\leq\bar{c}^{*}$.

On the other hand, let $\left[\tilde{P}_{1},\tilde{P}_{2},\tilde{P}_{3},\cdots\right]$
be an optimal solution to achieve $c^{*}$ in \eqref{eq: BCR1}. If
$\tilde{P_{i}}=0,\forall i>b-k$, then $\left[\bar{P}_{1},\bar{P}_{2},\bar{P}_{3},\cdots\bar{,P}_{b-k}\right]=\left[\tilde{P}_{1},\tilde{P}_{2},\tilde{P}_{3},\cdots\tilde{,P}_{b-k}\right]$
and $\bar{c}=c^{*}$ satisfy the constraints of \eqref{eq: BCR2},
which means $c^{*}\geq\bar{c}^{*}$.

If there exists $i>b-k$, such that $\tilde{P_{i}}>0$. Then we can
prove that $\left[\bar{P}_{1},\bar{P}_{2},\bar{P}_{3},\cdots\bar{,P}_{b-k}\right]=\left[\tilde{P}_{1},\tilde{P}_{2},\tilde{P}_{3},\cdots,\tilde{P}_{b-k}+\overset{\infty}{\underset{i=b-k+1}{\sum}}\tilde{P}_{i}\right]$
and $\bar{c}=c^{*}$ satisfy the constraints of \eqref{eq: BCR2}.
In fact, when $\bar{D}=D<b$, It is easy to verify that the coefficient
of $\tilde{P}_{i},\forall i>b-k$ is equal to the coefficient of $\tilde{P}_{b-k}$
in each constraint of \eqref{eq: BCR1}. Therefore, $\left[\bar{P}_{1},\bar{P}_{2},\bar{P}_{3},\cdots\bar{,P}_{b-k}\right]=\left[\tilde{P}_{1},\tilde{P}_{2},\tilde{P}_{3},\cdots,\tilde{P}_{b-k}+\overset{\infty}{\underset{i=b-k+1}{\sum}}\tilde{P}_{i}\right]$
and $\bar{c}=c^{*}$ satisfy the constraints of \eqref{eq: BCR2}.

Since when $D=2b-k-1$ in \eqref{eq: BCR1}, then we have

\begin{eqnarray}
 & \overset{2b-2k-1}{\underset{i=1}{\sum}}\left(b+i-1\right)P_{i}+\overset{\infty}{\underset{i=2b-2k}{\sum}}\left(2b-k-1\right)P_{i}\leq bc\label{eq:const11}
\end{eqnarray}

It is easy to verify that the coefficient of $\tilde{P}_{i},\forall i>b-k$
is equal to or greater than the coefficient of $\tilde{P}_{b-k}$.
Therefore, when $\bar{D}\geq b$, $\left[\bar{P}_{1},\bar{P}_{2},\bar{P}_{3},\cdots\bar{,P}_{b-k}\right]=\left[\tilde{P}_{1},\tilde{P}_{2},\tilde{P}_{3},\cdots,\tilde{P}_{b-k}+\overset{\infty}{\underset{i=b-k+1}{\sum}}\tilde{P}_{i}\right]$
and $\bar{c}=c^{*}$ still satisfy the constraints of \eqref{eq: BCR2}
due to \eqref{eq:const11}.

Hence, in both cases, $\left[\bar{P}_{1},\bar{P}_{2},\bar{P}_{3},\cdots\bar{,P}_{b-k}\right]=\left[\tilde{P}_{1},\tilde{P}_{2},\tilde{P}_{3},\cdots,\tilde{P}_{b-k}+\overset{\infty}{\underset{i=b-k+1}{\sum}}\tilde{P}_{i}\right]$
and $\bar{c}=c^{*}$ satisfy the constraints of \eqref{eq: BCR2},
we must have $c^{*}\geq\bar{c}^{*}$. Since we already proved that
$c^{*}\leq\bar{c}^{*}$, we must have $c^{*}=\bar{c}^{*}$.

Next, we are going to prove that an optimal solution $\bar{P}^{*}=\left[\bar{P_{1}}^{*},\bar{P_{2}}^{*},\bar{P_{3}}^{*},\cdots\bar{,P_{b-k}}^{*}\right]$
to \eqref{eq: BCR2} must satisfy that $\bar{P_{i}}^{*}>0,\forall i\leq b-k$.

First, if $\bar{P_{1}}^{*}=0$, let $j$ be the minimal $i$ such
that $\bar{P_{i}}^{*}>0$. Then it can be verified that the constraints
of \eqref{eq: BCR2} must hold as strict inequality for $\bar{D}\leq k+j-1$.

On the other hand, the coefficient of $\bar{P_{1}}^{*}$ must be less
than that of $\bar{P_{j}}^{*}$ in the constraints for $\bar{D}>k+j-1$.
Therefore, we can decrease $\bar{P_{j}}^{*}$ a little bit and increase
$\bar{P_{1}}^{*}$ a little bit such that all the constraints of \eqref{eq: BCR2}
have slackness, which means we can find a smaller $\bar{c}$ which
satisfies all the constraints. This is a contradiction that $\bar{P}^{*}=\left[\bar{P_{1}}^{*},\bar{P_{2}}^{*},\bar{P_{3}}^{*},\cdots\bar{,P_{b-k}}^{*}\right]$
is an optimal solution. Therefore, we must have $\bar{P_{1}}^{*}>0$.

Second, if there exists $h>1$ such that $\bar{P_{h}}^{*}=0$, then
we can decrease $\bar{P_{1}}^{*}$ a little bit and increase $\bar{P_{h}}^{*}$
a little bit. Since the coefficient of $\bar{P_{1}}^{*}$ must greater
than or equal to that of $\bar{P_{j}}^{*}$ in the constraints for
$\bar{D}\leq k+h-1$. On the other hand, when $\bar{D}\geq k+h$,
we want to compare the following constraints of $\bar{D}=k+1$ and
$\bar{D}\geq k+h$.

\begin{eqnarray*}
 & b\bar{P}_{1}+\left(k+1\right)\overset{b-k}{\underset{i=2}{\sum}}\bar{P}_{i}\leq\bar{c}\left(k+1\right),\\
 & \overset{h}{\underset{i=1}{\sum}}\left(b+i-1\right)\bar{P}_{i}+\overset{b-k}{\underset{i=h+1}{\sum}}\left(k+h\right)\bar{P}_{i}\leq\bar{c}\left(k+h\right),\\
 & \overset{h+1}{\underset{i=1}{\sum}}\left(b+i-1\right)\bar{P}_{i}+\overset{b-k}{\underset{i=h+2}{\sum}}\left(k+h\right)\bar{P}_{i}\leq\bar{c}\left(k+h+1\right)\\
 & \vdots
\end{eqnarray*}

When $\bar{D}\geq k+h$ , it is clear that the coefficient of $\bar{P_{h}}^{*}$
is at most $\left(h-1\right)$ greater than the coefficient of $\bar{P_{1}}^{*}$.
Therefore, when we decrease $\bar{P_{1}}^{*}$ a little bit and increase
$\bar{P_{h}}^{*}$ a little bit, the left side of those constraints
increase at most $\left(h-1\right)$ comparing to the case $\bar{D}=k+1$.
However, the right side increase at least $\left(h-1\right)\bar{c}$.
Hence, after we decreasing $\bar{P_{1}}^{*}$ a little bit and increasing
$\bar{P_{h}}^{*}$ a little bit, all the constraints of \eqref{eq: BCR2}
have slackness, which means we can find a smaller $\bar{c}$. This
is a contradiction that $\bar{P}^{*}=\left[\bar{P_{1}}^{*},\bar{P_{2}}^{*},\bar{P_{3}}^{*},\cdots\bar{,P_{b-k}}^{*}\right]$
is an optimal solution. Therefore, $\bar{P_{h}}^{*}>0,\forall h\in\left[2,b-k\right]$.
Up to now, we proved that an optimal solution $\bar{P}^{*}=\left[\bar{P_{1}}^{*},\bar{P_{2}}^{*},\bar{P_{3}}^{*},\cdots\bar{,P_{b-k}}^{*}\right]$
to \eqref{eq: BCR2} must satisfy that $\bar{P_{i}}^{*}>0,\forall i\leq b-k$.

Because \eqref{eq: BCR2} is a linear optimization problem and the
optimal value is not negative infinity, an optimal solution must be
a vertex of the polyhedron. Moreover, we have $\bar{P_{i}}^{*}>0,\forall i\leq b-k$.
Hence, the constraints $\bar{P_{i}}\geq0$ can not be active. On the
other hand, the dimension of variable vector is equal to the number
of the left independent constraints in \eqref{eq: BCR2}. Therefore,
an optimal solution must be the vertex that makes all the constraints
which are not $\bar{P_{i}}\geq0$ active, which means all the inequalities
must hold as equalities.

We can solve the linear equation system and get the minimal competitive
ratio and probability distribution:

\begin{eqnarray*}
 & c=\frac{\textrm{1}}{\textrm{1-\ensuremath{\left(\frac{b-k-1}{b-k}\right)^{b-k-1}\frac{b-k-1}{b}}}}\\
 & P_{b-k-i}=\frac{c}{b-k}\left(\frac{b-k-1}{b-k}\right)^{i},0\leq i<b-k-1\\
 & P_{1}=\left(\frac{b-k-1}{b-k}\right)^{b-k-1}\frac{k+1}{b}c,k<b
\end{eqnarray*}

Let $b$ go to infinity and $\frac{k}{b}=\alpha$, we have
\begin{eqnarray*}
 & c=\frac{e}{e-1+\alpha}
\end{eqnarray*}

This means the minimal competitive ratio for continuous time randomized
online algorithm is $c=\frac{e}{e-1+\alpha}$.

Therefore, we proved Theorem \ref{thm:online}.\end{IEEEproof}

\subsection{Proof of Corollary \ref{cormodified}\label{apx:proof_9}}
\begin{IEEEproof}
As we already showed before, under our last-empty-server-first job
dispatching strategy, each server actually serve the same set of job
both in online or offline situation. Moreover, the power consumption
of data center is minimal if each server runs off-line ski-rental
algorithm individually and independently in off-line situation. Therefore,
if each server runs online ski-rental algorithm individually and independently
in online situation, assume the competitive ratio of the online ski-rental
algorithm is $R$, then the total power consumption is at most the
minimal power consumption times $R$.

However, we must apply discrete time online ski-rental algorithm for
discrete-time fluid workload model because we chopped time into equal-length
slots. According to \cite{onlineski}, the competitive ratio of discrete
time online ski-rental algorithm is less than or equal to that of
continuous time online ski-rental problem. Therefore, our modified
deterministic and randomize online algorithms can retain competitive\textcolor{blue}{{}
}ratios $2-\alpha$, $\frac{e-\alpha}{e-1}$ and $\frac{e}{e-1+\alpha}$,
where $\alpha$ is the ratio of the number of time slots in future
window to the number of slots in critical interval $\vartriangle$.\end{IEEEproof}

\bibliographystyle{IEEEtran}
\bibliography{IEEEabrv,ref}

\begin{thebibliography}{10}
\providecommand{\url}[1]{#1}
\csname url@samestyle\endcsname
\providecommand{\newblock}{\relax}
\providecommand{\bibinfo}[2]{#2}
\providecommand{\BIBentrySTDinterwordspacing}{\spaceskip=0pt\relax}
\providecommand{\BIBentryALTinterwordstretchfactor}{4}
\providecommand{\BIBentryALTinterwordspacing}{\spaceskip=\fontdimen2\font plus
\BIBentryALTinterwordstretchfactor\fontdimen3\font minus
  \fontdimen4\font\relax}
\providecommand{\BIBforeignlanguage}[2]{{%
\expandafter\ifx\csname l@#1\endcsname\relax
\typeout{** WARNING: IEEEtran.bst: No hyphenation pattern has been}%
\typeout{** loaded for the language `#1'. Using the pattern for}%
\typeout{** the default language instead.}%
\else
\language=\csname l@#1\endcsname
\fi
#2}}
\providecommand{\BIBdecl}{\relax}
\BIBdecl

\bibitem{Koomey2008}
J.~G. Koomey, ``Worldwide electricity used in data centers,''
  \emph{Environmental Research Letters}, no.~3, 2008.

\bibitem{worldenergy07}
I.~E. Agency, ``World energy balances (2007 edition),'' 2007.

\bibitem{barroso2005price}
L.~Barroso, ``The price of performance,'' \emph{ACM Queue}, vol.~3, no.~7, pp.
  48--53, 2005.

\bibitem{ESPreport2007}
{U.S. Environmental Protection Agency}, ``Epa report on server and data center
  energy efficiency,'' \emph{ENERGY STAR Program}, 2007.

\bibitem{liu2011greening}
Z.~Liu, M.~Lin, A.~Wierman, S.~Low, and L.~Andrew, ``Greening geographical load
  balancing,'' in \emph{Proc. ACM SIGMETRICS}, 2011, pp. 233--244.

\bibitem{wendell2010donar}
P.~Wendell, J.~Jiang, M.~Freedman, and J.~Rexford, ``Donar: decentralized
  server selection for cloud services,'' in \emph{Proc. ACM SIGCOMM}, vol.~40,
  no.~4, 2010, pp. 231--242.

\bibitem{qureshi2009cutting}
A.~Qureshi, R.~Weber, H.~Balakrishnan, J.~Guttag, and B.~Maggs, ``Cutting the
  electric bill for internet-scale systems,'' in \emph{Proc. ACM SIGCOMM},
  2009, pp. 123--134.

\bibitem{urgaonkar2011optimal}
R.~Urgaonkar, B.~Urgaonkar, M.~Neely, and A.~Sivasubramaniam, ``Optimal power
  cost management using stored energy in data centers,'' in \emph{Proc. ACM
  SIGMETRICS}, 2011, pp. 221--232.

\bibitem{rasmussen113electrical}
N.~Rasmussen, ``Electrical efficiency modeling of data centers,''
  \emph{Technical Report White Paper}, vol. 113.

\bibitem{sharma2005balance}
R.~Sharma, C.~Bash, C.~Patel, R.~Friedrich, and J.~Chase, ``Balance of power:
  Dynamic thermal management for internet data centers,'' \emph{IEEE Internet
  Computing}, 2005.

\bibitem{raghavendra2008no}
R.~Raghavendra, P.~Ranganathan, V.~Talwar, Z.~Wang, and X.~Zhu, ``No power
  struggles: Coordinated multi-level power management for the data center,'' in
  \emph{ACM SIGARCH Computer Architecture News}, vol.~36, no.~1, 2008, pp.
  48--59.

\bibitem{chase2001managing}
J.~Chase, D.~Anderson, P.~Thakar, A.~Vahdat, and R.~Doyle, ``Managing energy
  and server resources in hosting centers,'' in \emph{Proc. ACM SOSP}, 2001.

\bibitem{pinheiro2001load}
E.~Pinheiro, R.~Bianchini, E.~Carrera, and T.~Heath, ``Load balancing and
  unbalancing for power and performance in cluster-based systems,'' in
  \emph{Workshop on Compilers and Operating Systems for Low Power}, 2001.

\bibitem{chen2008energy}
G.~Chen, W.~He, J.~Liu, S.~Nath, L.~Rigas, L.~Xiao, and F.~Zhao, ``Energy-aware
  server provisioning and load dispatching for connection-intensive internet
  services,'' in \emph{Proc. USENIX NSDI}, 2008.

\bibitem{krioukov2011napsac}
A.~Krioukov, P.~Mohan, S.~Alspaugh, L.~Keys, D.~Culler, and R.~Katz, ``Napsac:
  design and implementation of a power-proportional web cluster,'' \emph{ACM
  SIGCOMM Computer Communication Review}, vol.~41, no.~1, pp. 102--108, 2011.

\bibitem{fan2007power}
X.~Fan, W.~Weber, and L.~Barroso, ``Power provisioning for a warehouse-sized
  computer,'' in \emph{Proc. the 34th annual international symposium on
  Computer architecture}, 2007.

\bibitem{barroso2007case}
L.~Barroso and U.~Holzle, ``The case for energy-proportional computing,''
  \emph{IEEE Computer}, vol.~40, no.~12, pp. 33--37, 2007.

\bibitem{meisner2009powernap}
D.~Meisner, B.~Gold, and T.~Wenisch, ``Powernap: eliminating server idle
  power,'' \emph{ACM SIGPLAN Notices}, 2009.

\bibitem{qian2011server}
H.~Qian and D.~Medhi, ``Server operational cost optimization for cloud
  computing service providers over a time horizon,'' in \emph{Proceedings of
  the 11th USENIX conference on Hot topics in management of internet, cloud,
  and enterprise networks and services}, 2011, pp. 4--4.

\bibitem{lin2011dynamic}
M.~Lin, A.~Wierman, L.~Andrew, and E.~Thereska, ``Dynamic right-sizing for
  power-proportional data centers,'' \emph{Proc. IEEE INFOCOM, Shanghai,
  China}, pp. 10--15, 2011.

\bibitem{DataCenterEnergySurvey10}
A.~Beloglazov, R.~Buyya, Y.~C. Lee, and A.~Zomaya, ``A taxonomy and survey of
  energy-efficient data centers and cloud computing systems,'' \emph{Univ. of
  Melbourne, Tech. Rep. CLOUDS-TR-2010-3}, 2010.

\bibitem{gandhi2010optimality}
A.~Gandhi, V.~Gupta, M.~Harchol-Balter, and M.~Kozuch, ``Optimality analysis of
  energy-performance trade-off for server farm management,'' \emph{Performance
  Evaluation}, 2010.

\bibitem{doyle2003model}
R.~Doyle, J.~Chase, O.~Asad, W.~Jin, and A.~Vahdat, ``Model-based resource
  provisioning in a web service utility,'' in \emph{Proceedings of the 4th
  conference on USENIX Symposium on Internet Technologies and Systems}, 2003.

\bibitem{karlin1988competitive}
A.~Karlin, M.~Manasse, L.~Rudolph, and D.~Sleator, ``Competitive snoopy
  caching,'' \emph{Algorithmica}, vol.~3, no.~1, pp. 79--119, 1988.

\bibitem{karlin1994competitive}
A.~Karlin, M.~Manasse, L.~McGeoch, and S.~Owicki, ``Competitive randomized
  algorithms for nonuniform problems,'' \emph{Algorithmica}, vol.~11, no.~6,
  pp. 542--571, 1994.

\bibitem{bod2009statistical}
P.~Bod{\'\i}k, R.~Griffith, C.~Sutton, A.~Fox, M.~Jordan, and D.~Patterson,
  ``Statistical machine learning makes automatic control practical for internet
  datacenters,'' in \emph{Proceedings of the 2009 conference on Hot topics in
  cloud computing}.

\bibitem{narayanan2008write}
D.~Narayanan, A.~Donnelly, and A.~Rowstron, ``Write off-loading: Practical
  power management for enterprise storage,'' \emph{ACM Transactions on Storage
  (TOS)}, vol.~4, no.~3, p.~10, 2008.

\bibitem{kusic2009power}
D.~Kusic, J.~Kephart, J.~Hanson, N.~Kandasamy, and G.~Jiang, ``Power and
  performance management of virtualized computing environments via lookahead
  control,'' \emph{Cluster Computing}, vol.~12, no.~1, pp. 1--15, 2009.

\bibitem{onlineski}
\BIBentryALTinterwordspacing
C.~Mathieu, ``Online algorithms:ski rental.'' [Online]. Available:
  \url{http://www.cs.brown.edu/~claire/Talks/skirental.pdf}
\BIBentrySTDinterwordspacing

\end{thebibliography}

\end{document}